\documentclass[twocolumn,aps,showpacs,amsmath,amsfonts,amssymb,floatfix,pre]{revtex4}
\usepackage{graphicx}
\draft
\begin{document}

\title{Synchronized clusters in coupled map networks: Stability analysis}
\author {Sarika Jalan$^1$\footnote{e-mail: sarika@prl.ernet.in}, 
R. E. Amritkar$^1$\footnote{e-mail: amritkar@prl.ernet.in} and
Chin-Kun Hu$^2$\footnote{e-mail: huck@phys.sinica.edu.tw}}
\address{$^1$Physical Research Laboratory, Navrangpura, Ahmedabad 380 009,
India. \\ $^2$Institute of Physics, Academia Sinica, Nankang, Taipei 11529, Taiwan}

\begin{abstract}
We study self-organized (s-) and driven (d-) synchronization in coupled
map networks for some simple networks, namely two and three node
networks and their natural generalization to globally coupled and
complete bipartite networks. We use both linear stability
analysis and Lyapunov function approach for this study and determine
stability conditions for synchronization. 
We see that most of the features of coupled dynamics of
small networks with two or three nodes, are carried over to the larger networks
of the same type. 
The phase diagrams for the networks studied here have features very similar to
the different kinds of networks studied in Ref.~\cite{sarika-REA2}.
The analysis of the dynamics of the difference variable corresponding
to any two nodes shows that  when the two nodes are in
driven synchronization, all the coupling
terms cancel out whereas when they are in self-organized synchronization,
the direct coupling term between the
two nodes adds an extra decay term while the other 
couplings cancel out.
\end{abstract}


\maketitle
\section{Introduction}
Recently it has been observed that several complex systems have
underlying structures that are described by networks or graphs having
interesting properties \cite{Strogatz,rev-Barabasi}.
Many of these naturally occurring large and complex 
networks come under some universal classes and they can
be simulated with simple mathematical models, viz small-world 
networks \cite{Watts}, scale-free networks \cite{scalefree} etc. 
These models are based on simple physical considerations and have
attracted a lot of attention from physics 
community as they give simple algorithms to generate
graphs which resemble actual networks found in many 
diverse systems such as the nervous systems \cite{koch}, 
social groups \cite{social},
world wide web \cite{www}, metabolic networks \cite{metabolic},
food webs \cite{food} and citation networks \cite{citation}.

Several networks in real world consist of dynamical elements interacting
with each other and they have large number
of degrees of freedom. 
Synchronization in dynamical systems with many degrees of
freedom has attracted much attention in recent decades
\cite{book1-syn,book2-syn}.
In Refs.~\cite{sarika-REA1,sarika-REA2} two of us have 
presented detailed
analysis and numerical results of phase synchronization and cluster
formation in coupled maps
on different networks. Starting from random initial
conditions the asymptotic behaviour of these coupled map networks (CMNs)
has revealed that there are two different mechanisms
leading to two types of synchronized clusters. First there are clusters
with dominant intra-cluster couplings which are referred as {\it
self-organized (s-)} synchronization and secondly there are
clusters with dominant inter-cluster coupling which are referred as 
{\it driven (d-)}
synchronization \cite{sarika-REA1,sarika-REA2}.
The numerical studies
reveal several clusters with both types as well as clusters of mixed
type where both mechanisms contribute. There are also situations
where {\it ideal} clusters of both types are observed. An analysis of
simple networks
with two and three nodes indicates that
the self-organized behaviour has its origin in the
decay term arising due to intra-cluster couplings in the dynamics of
the difference variables while the driven behaviour has its origin in the
cancellation of the inter-cluster
couplings in the dynamics of the difference variables.

In the present paper we study the dynamics of some simple
networks analytically and numerically with a view to get a better
understanding of the two mechanisms of cluster formation discussed
above. Mainly we study the asymptotic stability of
s-  and d-synchronization in networks with small number
of nodes, i.e. two and three nodes, and extension of these small
networks to networks with large number of nodes.
As an example of large networks showing s-synchronized clusters
we take globally coupled maps \cite{kaneko-GCM} and for d-synchronized
clusters we take complete bipartite coupled maps \cite{bi-partite}.

The paper is organized as follows.
In Section II, we state the model and discuss general features of
the stability of synchronized states. Section III discusses two networks showing
s-synchronization, i.e. two nodes network and globally connected
network. Section IV considers two networks showing d-synchronization,
i.e. three nodes bipartite network and complete bipartite network. Section 
V concludes the paper.
 
\section{Model of a coupled map network (CMN)}
Consider a network of $N$ nodes and $N_c$ connections (or couplings)
between the nodes. Let each node of the
network be assigned a dynamical variable $x^i, i=1,2,\ldots,N$.
The dynamical evolution of coupled maps can be written as \cite{sarika-REA2}
\begin{equation}
x^{i}_{t + 1} = (1 - \epsilon) f( x^{i}_t ) + \frac{\epsilon}{k_i}
{\sum_{j=1}^{N} C_{ij} g( x^{j}_{t} ) }
\label{coupleddyn}
\end{equation}
where $x^{i}_t$ is the dynamical variable of the $i$-th 
node at
the $t$-th time step, $\epsilon$ is the coupling strength ($0 \leq \epsilon 
\leq 1$), $C$ is the
adjacency matrix with elements $C_{ij}$
taking values $1$ or $0$ depending upon whether $i$ and $j$ are
connected or not. $C$ is a symmetric matrix with diagonal 
elements zero. $k_i = \sum C_{ij}$ is the degree of node $i$.
The function $f(x)$ defines the local nonlinear map and
the function $g(x)$ defines the nature of coupling between the nodes.
For most of our results we use the logistic
map
\begin{equation}
f(x) = \mu x (1 - x)
\nonumber
\end{equation}
for illustration. We use two types of the
coupling functions,
\begin{subequations}
\begin{eqnarray}
g(x) &=& x \\ 
g(x) &=& f(x)
\end{eqnarray}
\end{subequations}
We refer to the first type of coupling as linear and the latter as nonlinear.

{\it Synchronization}:
Synchronization of two dynamical variables or systems is indicated by the
appearance of some relation between
functionals of two processes due to interaction \cite{book1-syn,book2-syn,sarika-REA1,sarika-REA2}) 
In this paper we will mostly concentrate on 
exact synchronization, where the values of the dynamical variables 
associated with nodes are equal.
We can see that the fully synchronized state of a network, $x^1_t =
x^2_t = \cdots = x^N=x_t$, is a solution to Eq.~(\ref{coupleddyn}). This fully
synchronized state
lies along a one-dimensional diagonal in the $N$ dimensional phase
space of the dynamical variables.
We define a synchronized cluster as a cluster of nodes in which
all pairs of nodes are synchronized. 
As stated in the introduction, we can identify two different mechanism of
cluster formation in CMNs. Self-organized synchronization
which leads to clusters with dominant intra-cluster couplings and 
d-synchronization which leads to clusters with dominant inter-cluster couplings
\cite{sarika-REA1}.

In this paper we study synchronization in some simple networks. 
We use two types of analysis to determine the stability
of synchronized state.  
First is the linear stability analysis 
\cite{GCM-stab1,REA-gade,GCM-stab2,GCM-stab3,CML-stability1,CMN-stability}
and the second is
Lyapunov functional \cite{book-NLD,Lya-fun,CMN-stability}. We now
briefly discuss these two methods.

{\it Linear Stability Analysis}:
The evolution of tangent vector $\delta z_t = [\delta x_t^1, \delta x_t^2
\dots, \delta x_t^N]^T$, along a trajectory can be written as,
\begin{equation}
\delta z_{t+1} = J_t \delta z_t
\nonumber
\end{equation}
where $J_t$ is the Jacobian matrix at time $t$, $(J_t)_{ij} = \partial 
x^i_{t+1} /\partial x^j_t$. If $J_t$ is a diagonal
matrix or the similarity transformation which diagonalizes $J_t$ is
independent of $t$, then the Lyapunov exponents can be written in
terms of the eigenvalues of $J_t$ as follows,
\begin{equation}
\lambda_i = \lim_{\tau \rightarrow \infty} \frac{1}{\tau}
\sum_{t=1}^\tau \ln \left| \Lambda_i(t)\right |
\end{equation}
where $\Lambda_i(t)$ is the $i$-th eigenvalue of Jacobian matrix
at time $t$. If $J_t$ does not satisfy the above mentioned
conditions then it is necessary to
consider product of Jacobian matrices to obtain Lyapunov
exponents. 

To study the stability of synchronized state it is sufficient to consider
transverse Lyapunov exponents which characterize the behavior of
infinitesimal vectors transversal to the synchronized manifold,
and these determine the stability of a synchronized state \cite{book-NLD}.
 If all 
the transverse Lyapunov exponents are negative then the synchronized
state is stable.

{\it Global Stability Analysis}:
Condition for global stability can be derived using the 
Lyapunov functional methods \cite{book-NLD,Lya-fun}. 
Global stability in a neighborhood of an equilibrium (stable) point is
confirmed if there exist 
a positive definite function defined in that neighborhood, whose
total time derivative is negative semi-definite.
 
To get the conditions for the global stability of
synchronization of two trajectories $x^i_t$ and $x^j_t$,
we define Lyapunov function as,
\begin{equation}
V^{ij}_t = (x^i_t - x^j_t)^2
\label{Lya-fun}
\end{equation}
Clearly $V_{ij}(t) \ge 0$ and the equality holds only when the nodes
$i$ and $j$ are exactly synchronized.
For the asymptotic global stability of the synchronized state,
 Lyapunov function
should satisfy the following condition in the region of stability,  
\begin{equation}
V_{t+1} <  V_t   \nonumber
\end{equation}
This condition can also be written as,
\begin{equation}
\frac{V_{t+1}}{V_t} < 1.
\label{cond-syn} 
\end{equation}

\section{Self-organized Synchronization}
We first consider the simplest and smallest network showing
s-cluster,
i.e synchronization of two coupled nodes which is obviously of self-organized
type. As a generalization of two nodes network to large networks, we
study the s-synchronization
in globally coupled maps.

\subsection{Coupled network with $N = 2$}
We begin by taking the simplest case where number of nodes is two
and these two nodes are coupled with each other \cite{bi-partite}.
The dynamics of the two nodes can be rewritten as (Eq.~(\ref{coupleddyn})),
\begin{eqnarray}
x^1_{t+1} &=& (1 - \epsilon) f(x^1_t) + \epsilon g(x_t^2) \nonumber  \\
x^2_{t+1} &=& (1 - \epsilon) f(x^2_t) + \epsilon g(x_t^1)
\label{model-2}
\end{eqnarray}

\subsubsection{Linear stability analysis}
Following
Ref.~ \cite{book2-syn}, we first define addition and difference
variables as follows,
\begin{eqnarray}
s_t &=& \frac{x^1_t + x^2_t}{2} \nonumber \\
d_t &=& \frac{x^1_t - x^2_t}{2}
\label{def-add-diff}
\end{eqnarray}
Dynamical evolution for these newly defined variables is given by,
\begin{subequations}
\begin{eqnarray}
s_{t+1} &=& \frac{1-\epsilon}{2} [ f( s_t+d_t )+ f( s_t-d_t ) ] 
\nonumber \\
 & & + \frac{\epsilon}{2} [ g( s_t+d_t )+ g( s_t-d_t ) ] \nonumber \\
d_{t+1} &=& \frac{1-\epsilon}{2} [ f(s_t+d_t)-f(s_t-d_t) ] 
  \nonumber\\
 & & - \frac{\epsilon}{2} [ g(s_t+d_t)-g(s_t-d_t) ] 
\label{add-diff-2}
\end{eqnarray}
\end{subequations}
For synchronous orbits to be observed, the fully
synchronized state $d_t=0$ i.e. $x^1_t=x^2_t=x_t=s_t$, should be a
stable attractor.
The Jacobian matrix for the synchronized state is,
\begin{equation}
J_t = \left( \begin{array}{ccc} 
(1-\epsilon)f^\prime(x_t) + \epsilon g^\prime(x_t) & 0\\ 
0 & (1-\epsilon)f^\prime(x_t) - \epsilon g^\prime(x_t) 
\end{array}  \right) \nonumber
\end{equation}
where the prime indicates the derivative of the function.
The above Jacobian is a diagonal matrix and Lyapunov exponents can be easily 
written in terms of eigenvalues of product of such Jacobian matrices
calculated at different time. 
The two Lyapunov exponents are
\begin{eqnarray}
\lambda_{s,d} & = & \lim_{\tau \rightarrow \infty} 
\frac{1}{\tau} {\sum_{t=1}^\tau \ln|(1-\epsilon)f^\prime(x_{t}) \pm 
\epsilon g^\prime(x_t)| }
\label{lya-2}
\end{eqnarray}
The synchronous orbits are stable if Lyapunov exponent corresponding
to the difference variable $d_t$, i.e. $\lambda_d$ or the transverse
Lyapunov exponent, is negative. 
If the other Lyapunov exponent $\lambda_s$ is positive then the
synchronous orbits are chaotic while if it is negative then they
are periodic.
  
\noindent{\it Coupling function $g(x) = f(x)$}:
Two coupled maps with $g(x)=f(x)$ type of coupling are studied extensively 
in the literature both 
analytically and numerically \cite{book2-syn,driven-T1S2}.
For the synchronized state Lyapunov exponent $\lambda_s$
is nothing but the Lyapunov exponent for uncoupled
logistic map ($\lambda_u$) and the other Lyapunov exponent can be written
in terms of the $\lambda_u$ and from Eq.~(\ref{lya-2}) we get
\begin{subequations}
\begin{eqnarray}
\lambda_s & = &  \lambda_u = \lim_{\tau \rightarrow \infty} 
\frac{1}{\tau} {\sum_{t=1}^\tau \ln|f^\prime (x_{t})| }  \\
\lambda_d &=& \ln |1-2\epsilon| + \lambda_u
\end{eqnarray}
\end{subequations}
The synchronous orbits are stable if Lyapunov exponent corresponding
to the difference variable $d_t$ is negative,
i.e. $\lambda_d < 0$. Thus the range of stability of the synchronized
state is given by
\begin{equation}
\frac{1-e^{-\lambda_u}}{2} < \epsilon < 0.5 \frac{1+e^{-\lambda_u}}{2}
\end{equation} 
For logistic map with $\mu=4$, this gives $ 0.25 < \epsilon < 0.75 $
as the range for the stability of the synchronized state.

\noindent{\it Coupling function $g(x) = x$}:
For $g(x)=x$ type of coupling, numerical results show that
as the coupled nodes evolve, dynamics shows different types of synchronized 
and periodic behaviors depending upon the coupling strength $\epsilon$
and the parameter of the map $f$.
First let us start with the general case where coupled dynamics
lies on a synchronized attractor.
Using Eq.~(\ref{lya-2}), Lyapunov exponents can be
easily written as
\begin{equation}
\lambda_{s,d} = \lim_{\tau \to \infty} \frac{1}{\tau} {\sum_{t=1}^\tau 
\ln|(1-\epsilon)f^{\prime}(s_t) \pm \epsilon| }  \nonumber
\end{equation}
For stable synchronous orbits Lyapunov exponent corresponding to
the difference variable, i.e. $\lambda_d$, should be negative.
Now we consider some special cases when coupled dynamics lies on
periodic or fixed point attractor. 

Here, as well as for other networks 
considered in this paper, we will restrict ourselves to
period two orbits. Higher periodic orbits exist but are 
difficult to treat analytically. Also major features of phase diagram
are understood by using fixed point and period two orbits.

{\it Case I. Synchronization to period two orbit}:
Consider the special case where the solution of 
Eq.~(\ref{model-2}) or Eq.~(\ref{add-diff-2}) is a periodic orbit of 
period two,
with the difference variable given by $d_{t+1}=d_t=0$ and 
addition variable given by $s_{t+2} = s_t = s_1, \, s_{t+3}=s_{t+1} = s_2$.
Eigenvalues for the product matrix $J_1 J_2$, where $J_1$ and $J2$
are Jacobian matrices for consecutive time steps, are given by
\begin{equation}
\Lambda_{1,2} = (1-\epsilon)^2 f^{\prime}_1 f^{\prime}_2 + \epsilon^2 
\underline{+} (1-\epsilon) \epsilon (f^{\prime}_1 + f^{\prime}_2)
\label{Lambda1-2-x}
\end{equation}
where $f^{\prime}_1$ and $f^{\prime}_2$ are derivatives of $f$
at the two periodic points $(s_1,d=0)$ and $(s_2,d=0)$ respectively.
The $\epsilon$ range for which the dynamical evolution 
gives a stable periodic orbit, is obtained when the modulus of the 
eigenvalues for matrix $J_1 J_2$ are less than one. 
For local dynamics given by logistic map $f(x)=\mu x (1-x)$,
the two periodic points are,
\begin{equation}
s_{1,2} = \frac{1+\epsilon + \mu(1-\epsilon) \pm 
\sqrt{\epsilon(\epsilon-2)(\mu-1)^2 -3 - 2\mu + \mu^2}}{2\mu(1-\epsilon)}
\label{mu-case1-2-x}
\end{equation}
For $\mu=4$,
\begin{equation}
s_{1,2} = \frac{5-3 e + \sqrt{5-18 e + 9 e^2}}{8(1 - e)}
\label{sol-case1-2-x}
\end{equation}
which gives the coupling strength range $0.18.. < \epsilon < 0.24..$ for which
the periodic orbit,  $(s_1,d=0)$ and $(s_2,d=0)$, is stable.

{\it Case II. Period two orbit}:
There is a range of $\epsilon$ values that give the following
stable period two behaviour,
\begin{equation}
\begin{array}{rccccl}
x^1_{t} &=& x^2_{t+1} = x^1_{t+2} &=& X_1^p, \nonumber \\
x^2_{t} &=& x^1_{t+1} = x^2_{t+2} &=& X_2^p
\label{case2-2-x}
\end{array}
\end{equation}
Lyapunov exponents for this periodic state can be found from
the eigenvalues of the product of Jacobians at the two periodic
points. Jacobian matrix at $(X_1^p,X_2^p)$ is given by,
\begin{equation}
J_1 = \left( \begin{array}{ccc}
(1-\epsilon)f^{\prime}_1 & \epsilon\\
\epsilon & (1-\epsilon)f^{\prime}_2
\end{array} \right) 
\label{jacobian-case2-2-x}
\end{equation}
where $f^{\prime}_1$ and $f^{\prime}_2$ are the derivative of
$f$ at $X_1^p$ and $X_2^p$ respectively.
Eigenvalues of the product matrix $J_1 J_2$ are,
\begin{equation}
\Lambda_{1,2} = \left( \epsilon  
\pm (1-\epsilon) \sqrt{f^{\prime}_1f^{\prime}_t} \right)^2
\label{Lambda-case2-2-x}
\end{equation}
If $f^\prime_1 f^\prime_2 <0 $ which is the interesting case, then
the condition for the stability of the periodic orbit become
\begin{equation}
\frac{f^\prime_1 f^\prime_2 - 1}{f^\prime_1 f^\prime_2 +1} < \epsilon <1.
\label{range-2-case2-x}
\end{equation} 
For $f(x)=\mu x (1-x)$, $X_1^p$ and $X_2^p$ which satisfy 
Eq.~(\ref{case2-2-x}) are given by
\begin{equation}
X_{1,2}^p = \frac{ (1+\frac{1}{\mu}) \pm \sqrt{(1+\frac{1}{\mu})^2 - 
\frac{4}{\mu}(1+\frac{1}{\mu})}}{2}
\label{sol-case2-2-x}
\end{equation}
For stable periodic orbits modulus of both eigenvalues
$\Lambda_1$ and $\Lambda_2$ are less than one which gives the coupling
strength range for stability as,
\begin{equation}
1 - \frac{2}{\mu^2-2\mu-3} < \epsilon \leq 1
\label{range-case2-2-x}
\end{equation}
For $\mu=4$, we get coupling strength range $(0.6 < \epsilon < 1)$ 
for which periodic orbit as given in Eq.(\ref{case2-2-x}) is stable and 
coupled dynamics lies on a periodic attractor.

\subsubsection{Lyapunov function analysis}

From Eqs.~(\ref{Lya-fun}) and~(\ref{model-2}),
Lyapunov function for two nodes is written as
\begin{eqnarray}
V^{12}_{t+1} & = & V_{t+1} \nonumber \\ 
 & = & \left[ (1- \epsilon)(f(x^1_t)-f(x^2_t)) -
 \epsilon (g(x^1_t) - g(x^2_t)) \right]^2 
\label{Lya-fun-2} 
\end{eqnarray}
Using Taylor expansion of $f(x^1_t)$ and $g(x^1_t)$ about $x^2_t$, 
we get Lyapunov function at time $t+1$ as,
\begin{eqnarray}
V_{t+1} & = & V_t \left[ (1-\epsilon)f^{\prime}(x^2_t) - 
\epsilon g^\prime(x^2_t) \right. \nonumber \\ 
 & & + \frac{x^1_t - x^2_t}{2} ((1-\epsilon) f^{\prime\prime}(x^2_t) -
\epsilon) g^{\prime\prime}(x^2_t) ) \nonumber \\ 
  + & & \left. {\cal O}[(x^1_t - x^2_t)^2 \right]^2 
\label{lyafun-2}
\end{eqnarray}
If the expression in the square bracket on the RHS is less than one then
the synchronized state is stable.

For the nonlinear coupling function $g(x)=f(x)$ the expression
 (\ref{lyafun-2}) for Lyapunov function simplifies and we get
\begin{equation}
\frac{V_{t+1}}{V_t} = (1-2\epsilon)^2 [f^{\prime}(x^2_t)  
\frac{x^1_t - x^2_t}{2} f^{\prime\prime}(x^2_t) +
{\cal O}[(x^1_t - x^2_t)^2]^2 
\end{equation}
If the expression in the square bracket on the RHS is bounded then
there will always some range of $\epsilon$ values around 0.5 for which 
the synchronized state will be stable.
For logistic map  $f(x) = \mu x(1-x)$, we get,
\begin{equation}
V_{t+1} = V_t (1 - 2 \epsilon)^2 \mu^2 [ 1 - (x^1_t + x^2_t) ]^2 \nonumber
\end{equation}
Using $0 \leq x^1_t + x^2_t \leq 2$, 
we get the following range of $\epsilon$ values for which the synchronization
condition given by Eq.~(\ref{cond-syn}) is satisfied,
\begin{equation}
\frac{1}{2}\left(1-\frac{1}{\mu}\right)  
< \epsilon < \frac{1}{2}\left(1+\frac{1}{\mu}\right)  
\end{equation}
For $\mu=4$, it gives the coupling strength range 
$0.325 < \epsilon < 0.675$.
However, a better $\epsilon$ range can be obtained by putting
more realistic bounds for $x^1_t+x^2_t$ as
\begin{equation}
\frac{1}{2}\left(1-\frac{1}{\mu X}\right) < \epsilon < \frac{1}{2}
\left(1+\frac{1}{\mu X}\right) 
\end{equation}
where $X = \sup_t (|1-x^1_t-x^2_t|)$.

\subsection{Globally coupled networks}
Globally coupled networks have all pairs of nodes connected
to each other i.e. $N_c = N(N-1)/2$ where $N$ is the number of nodes.
For such global coupling we write our dynamical model as
\begin{equation}
x^i_{t+1} = (1-\epsilon) f(x^i_t) + \frac{\epsilon}{N-1} 
{\sum^N_{j \neq i,j=1} g(x^j_t) }
\label{GCM}
\end{equation}
Let the state $x^1_t = x^2_t = \cdots = x^N_t = x_t$ be the fully synchronized
state. We now consider the stability of this state. 

\subsubsection{Linear Stability Analysis}

Jacobian matrix at time $t$ for the fully synchronized state is  
\begin{equation}
J_t = \left( \begin{array}{cccc}
(1-\epsilon) f^{\prime}_t & \frac{\epsilon}{N-1}g^\prime_t &
       \dots & \frac{\epsilon}{N-1}g^\prime_t \\
\frac{\epsilon}{N-1}g^\prime_t & (1-\epsilon) f^{\prime}_t & 
        \dots &\frac{\epsilon}{N-1}g^\prime_t \\
\vdots & \vdots & \vdots & \vdots\\
        \frac{\epsilon}{N-1}g^\prime_t & \frac{\epsilon}{N-1}g^\prime_t &
        \dots & (1-\epsilon) f^{\prime}_t
\end{array} \right) \nonumber
\end{equation} 
where $f^{\prime}_t$ and $g^\prime_t$ are the derivative at the synchronous
 value $x_t$. Eigenvectors of the above Jacobian matrix are,
\begin{equation}
E_m = \left( exp(2 \pi i \frac{m}{N}), exp(4 \pi i \frac{m}{N}) \dots,
exp(2 N \pi i \frac{m}{N}) \right )^T \nonumber
\end{equation}
where $m= 0,1, \dots, N-1$ and $T$ denotes the transpose. From these 
eigenvectors we find that $J_t$ has an eigenvalue  
$(1-\epsilon) f^{\prime}_t + \epsilon g^\prime_t$ and $(N-1)$-fold degenerate 
eigenvalues
$(1-\epsilon)f^{\prime}_t - \frac{\epsilon}{N-1} g^\prime_t$. Lyapunov exponents
can be written in terms of eigenvalues of Jacobian matrix as,
\begin{subequations}
\begin{eqnarray}
\lambda_1 &=& \lim_{\tau \rightarrow \infty} \frac{1}{\tau} {\sum_{t=1}^\tau
{\ln \left| (1-\epsilon) f^{\prime}_{t} + \epsilon g^\prime_t \right| } } 
\label{lambda1-gcm} \\
\lambda_2 & = & \lambda_3 \dots =\lambda_N \nonumber \\
   &=& \lim_{\tau \rightarrow \infty} 
\frac{1}{\tau} \sum_{t =1}^\tau \ln 
\left | (1-\epsilon)f^{\prime}_{t} 
- \frac{\epsilon}{N-1}g^\prime_t  \right| 
\label{lambda2-gcm}
\end{eqnarray}
\end{subequations}
Lyapunov exponents $\lambda_2, \lambda_3 \dots \lambda_n$ are transverse
Lyapunov exponents since they characterize the behaviour of infinitesimal 
vectors transverse to the synchronization manifold.
For stability of the synchronous orbits all transverse Lyapunov exponents
should be negative.

{\it Coupling Function $g(x) = f(x)$}:
Globally coupled maps are studied extensively with $g(x)=f(x)$
type of coupling and linear stability analysis is done to decide the
stability of fully synchronized solution 
\cite{GCM-stab1,GCM-stab2,GCM-stab3}.
Lyapunov exponents are given by, 
\begin{subequations}
\begin{eqnarray}
\lambda_1 & = &  \lambda_u = \lim_{\tau \rightarrow \infty} 
{\sum_{t=1}^\tau \ln|f^\prime (x_{t})| }  \\
\lambda_2 & = & \lambda_3 = \dots = \lambda_N = \lambda_u + 
\ln(1-\frac{N}{N-1}\epsilon) 
\end{eqnarray}
\end{subequations}
Except $\lambda_u$ other Lyapunov exponents are transverse to the
synchronization manifold.
The critical value of coupling strength $\epsilon_c$ beyond which the
synchronized state with all nodes synchronized with each other is
stable, is given by,
\begin{equation}
\epsilon_c = \frac{N-1}{N} (1 - e^{-\lambda_u})
\end{equation}
For large $N$ with $f(x)= 4 x (1-x)$, we have $\lambda_u = \ln 2$.
So from above expression we
get $\epsilon_c = 0.5$.

{\it Coupling function $g(x) = x$}:
From the expressions (\ref{lambda1-gcm}) and~(\ref{lambda2-gcm}), it
is difficult to determine when the synchronous orbits are
stable. Rather it is easier to
determine the stability of the synchronous orbits using Lyapunov function
which we will consider in the next subsection.
Here, we consider some special cases when coupled dynamics of the fully
synchronized state lies on periodic or fixed point attractors.

{\it Case I. Synchronization to fixed point}: First we consider the
fixed point $X^*=x_t$ of the fully
synchronized state. The fixed point is given by $X^* = f(X^*)$.
The conditions for the
stability of the synchronous fixed point are
$\ln|(1-\epsilon)f^\prime+\epsilon| <0$ and 
$\ln|(1-\epsilon)f^\prime-\frac{\epsilon}{N-1}| <0$. 
where $f^\prime$ is the derivative of $f$ at $X^*$. For $f^\prime <0$
which is the interesting case, the fixed point is easily found to be
stable in the range
\begin{equation}
\frac{|f^\prime|-1}{|f^\prime|-\frac{1}{N-1}} < \epsilon < 1.
\end{equation}
For $N=2$ this solution is not stable and the range of stability
increases with $N$. This is a surprising result since with
increasing $N$ the number of transverse eigenvectors along which the
synchronized solution can become unstable also increases. This feature
of increasing range of stability with $N$ is more general and will be
noticed for other
solutions also as will be discussed in subsequent analysis.  
For large $N$ synchronous fixed point is stable for
\begin{equation}
\frac{|f^\prime|-1}{|f^\prime|} < \epsilon < 1
\end{equation}
For $f(x)$ given by logistic map, $X*=1-1/\mu, f^\prime(X^*) = 2 - \mu$ 
and range of stability of the synchronous fixed point is given by
\begin{equation}
\frac{\mu-3}{\mu-2-\frac{1}{N-1}} < \epsilon < 1.
\end{equation} \\

{\it Case II. Synchronization to period two orbit}: We now consider
synchronous period two solution.
 The Lyapunov exponents can be obtained using
Eqs.~(\ref{lambda1-gcm}) and~(\ref{lambda2-gcm}). The conditions
for the stability of the period two solution are
\begin{subequations}
\begin{eqnarray}
\ln \left| \left((1-\epsilon)f^{\prime}_1 
+\epsilon\right) \left( (1-\epsilon)f^{\prime}_2 
+\epsilon\right)  \right|  & < & 0 \\
\ln \left| \left((1-\epsilon)f^{\prime}_1 
- \frac{\epsilon}{N-1}\right) \left( (1-\epsilon)f^{\prime}_2 
- \frac{\epsilon}{N-1}\right)  \right|  & < & 0
\end{eqnarray}
\end{subequations}
For $f(x)$ given by logistic map, the period two synchronized solution
is the same as given by Eqs.~(\ref{mu-case1-2-x})
and~(\ref{sol-case1-2-x}).
For $\mu=4$, the range of stability of this solution is
\begin{equation}
1-\frac{\sqrt{54}}{9} < \epsilon < \frac{10 + \frac{1}{N-1} -
\sqrt{60+\frac{30}{N-1} + \frac{6}{(N-1)^2} }} {8-\frac{2}{N-1} - 
\frac{1}{(N-1)^2} }
\label{range-case2-gcm-x}
\end{equation}
For $N=2$, this gives the $\epsilon$ range (0.18..,0.24..) for the
stability of period two solution as noted in Case1 of
two node case with $g(x)=x$. For large $N$, this range of stability
for $\epsilon$ values expands to (0.18..,0.28..) \cite{note1}.  

\subsection{Lyapunov Function Analysis}
From Eqs. (\ref{Lya-fun}) and (\ref{GCM}), we write Lyapunov function for
any two nodes of globally coupled network as,
\begin{equation}
V^{i,j}_{t+1} = \left[ (1-\epsilon)\left( f(x^i_t)-f(x^j_t) \right) -
\frac{\epsilon}{N-1}\left( g(x^i_t) - g(x^j_t) \right) \right]^2
\nonumber
\end{equation}
Performing Taylor expansion around $x^j_t$, we get
\begin{eqnarray}
\frac{V_{t+1}}{V_t} & = & \left [(1-\epsilon) f^{\prime}(x^j_t) -
    \frac{\epsilon}{N-1} g^\prime(x^j_t)  \right. \nonumber \\
    & & + \frac{x^i_t - x^j_t}{2} \left((1-\epsilon) f^{\prime\prime}(x^j_t) - 
    \frac{\epsilon}{N-1} g^{\prime\prime}(x^j_t) \right) \nonumber \\
    & & + \left. {\cal O}[(x^1_t - x^2_t)^2 \right]^2 
\label{lyafun-global}
\end{eqnarray}

{\it Coupling function $g(x)=f(x)$}: 
In this case the expression (\ref{lyafun-global}) simplifies to
\begin{eqnarray}
\frac{V_{t+1}}{V_t} & = & \left(1 - \frac{N}{N-1} \epsilon \right)^2 
\left [f^{\prime}(x^j_t) + \frac{x^i_t - x^j_t}{2} 
f^{\prime\prime}(x^j_t) \right. \nonumber \\
  & & + \left. {\cal O}[(x^1_t - x^2_t)^2 \right]^2 \nonumber 
\end{eqnarray}
If the expression in the square bracket on the RHS is bounded then for
large $N$ there will be a critical value of $\epsilon$ 
beyond which the condition (\ref{cond-syn}) will be
satisfied and the globally synchronized state will be stable.
For $f(x)=\mu x (1-x)$ and using  $(0 \leq x^i_t+x^j_t \leq 2)$, we
get the following range of coupling strength values for which 
the globally synchronized state is stable.
\begin{equation}
\frac{N-1}{N} \left( 1 - \frac{1}{\mu} \right) 
< \epsilon \leq 1 < \frac{N-1}{N} \left( 1 + \frac{1}{\mu} \right)  
\end{equation}
For $\mu=4$ and for very large $N$, coupling strength range
is $ 0.75 < \epsilon < 1$.
A better $\epsilon$ range is obtained by putting
more realistic bounds as $X^- \le x^i_t+x^j_t  \le X^+$, 
which gives the range of stability as,
\begin{equation}
\frac{N-1}{N}\left(1-\frac{1}{\mu A}\right)
< \epsilon \leq 1 
\label{range-gcm-lyafun-fx}
\end{equation}
where $A = \max(|1-X^+|,|1-X^-|)$.

{\it Coupling function $g(x)=x$}:
In this case, for logistic map, the expression (\ref{lyafun-global})
 simplifies to
\begin{equation}
\frac{V^{ij}_{t+1}}{V_t}= \left[ (1-\epsilon)(1-( x^i_t+x^j_t )) - 
\frac{\epsilon}{N-1} \right]^2   \nonumber
\end{equation}
Since $0 \leq x^i_t+x^j_t \leq 2$ 
we get a range of $\epsilon$ values for which the globally synchronized state
is stable (Eq.~(\ref{cond-syn}))
\begin{equation}
\frac{\mu - 1}{\mu - \frac{1}{N-1}} < \epsilon < 
\frac{\mu+1}{\mu - \frac{1}{N-1}} 
\label{eps-range-GCM-x}
\end{equation}
which for large $N$ reduces to
\begin{equation}
\frac{\mu-1}{\mu} < \epsilon < \frac{\mu+1}{\mu}
\end{equation}
For $\mu=4$, we get the coupling strength range $0.75 < \epsilon \leq 1$
for which the globally synchronized state is stable.
We also note that for $N=2$ the
condition~(\ref{eps-range-GCM-x}) for synchronization is not satisfied 
for any value of $\epsilon \leq 1$. 

\section{Driven Synchronization}
Driven synchronization leads to clusters with dominant inter-cluster
couplings. For the ideal d-clusters, there are only inter-cluster
connections with
no connections between the constituents of the same cluster \cite{sarika-REA1}.
A complete bipartite network consists of two sets of nodes with each node of one
set connected
with all the nodes of the other set. Clearly this type of network is
an ideal example for studying d-synchronization. 
We take a bipartite network consisting of two sets of nodes,
$N=N1+N2$, with each node of $N1$ connected to every node of $N2$ and 
there are no connections between the nodes of the same set \cite{bi-partite}.  
We study dynamics of coupled maps on such type of bipartite network
and determine the stability criteria for formation of
d-synchronized clusters.
Our model for coupled complete bipartite network can be written as,  
\begin{eqnarray}
x^i_{t+1}& = & 
(1-\epsilon) f(x^i_t) + \frac{\epsilon}{N2} {\sum^N_{j=N1+1} g(x^j_t) }, 
\nonumber \\ & & \; \; \; \; \; {\rm for} \, i = 1,\cdots, N1 
 \nonumber \\
& = & (1-\epsilon) f(x^i_t) + \frac{\epsilon}{N1} {\sum^{N1}_{j=1} g(x^j_t)},
\nonumber \\ & & \; \; \; \; \; {\rm for} \, i =  N1+1,\cdots, N.
\label{model-driven}
\end{eqnarray}
where all terms are having the same meaning as defined for 
Eq.~(\ref{coupleddyn}).

\subsection{Network with three nodes}
First we take a simple network of three nodes with $N1=2,
N2=1$ which is the
smallest possible network to show the behaviour displayed by
Eq.~(\ref{model-driven}). The evolution equations can be written as
\begin{eqnarray}
\begin{array}{rcl}
x^1_{t+1} &=& (1-\epsilon) f(x^1_t) + \epsilon g(x^3_t)\\
x^2_{t+1} &=& (1-\epsilon) f(x^2_t) + \epsilon g(x^3_t)\\
x^3_{t+1} &=& (1-\epsilon)f(x^3_t) + \frac{\epsilon}{2}( g(x^1_t) + g(x^2_t) )
\end{array}
\label{model-3}
\end{eqnarray}
Here, the d-synchronized state corresponds to the nodes 1 and 2 being
synchronized with each other but not with node 3.

\subsubsection{Linear Stability Analysis}
Here, we use 
addition and difference variables $s_t$ and $d_t$  
as defined in Eq.~(\ref{def-add-diff}) for the first two nodes.
Thus Eqs.~(\ref{model-3}) can be rewritten as,
\begin{subequations}
\begin{eqnarray}
s_{t+1} &=& \frac{1-\epsilon}{2} [ f(s_t+d_t) + f(s_t - d_t)] +
\epsilon g(x_t^3)\\
d_{t+1} &=& \frac{1-\epsilon}{2} [f(s_t+d_t) - f(s_t - d_t)]
\label{diff-3} \\
x^3_{t+1} &=& (1-\epsilon) f(x^3_t) + \frac{\epsilon}{2}[g(s_t+d_t) +
g(s_t-d_t)
\end{eqnarray}
\end{subequations}
Jacobian matrix for the d-synchronized state $(s_t, d_t=0, x^3_t)$
is given by,
\begin{equation}
J_t = \left( \begin{array}{cccc}
(1-\epsilon) f^{\prime}_1 & 0 & \epsilon g_2^{\prime}\\
0 & (1-\epsilon) f^{\prime}_1 & 0\\
\epsilon g^{\prime}_1 & 0 & (1-\epsilon) f^{\prime}_2 \\
\end{array} \right )
\label{jacobian-3}
\end{equation}
$f^{\prime}_1$ and $g^{\prime}_1$ are derivatives at
 $s_t=x_t=x^1_t=x^2_t$ and $f^{\prime}_2$ and $g^{\prime}_2$ are 
derivatives at $(x^3_t)$.
Lyapunov exponent corresponding to the difference variable is found as,
\begin{eqnarray}
\lambda_d & = &\ln(1-\epsilon) + \lambda_c
\label{lambda-3-fx}
\end{eqnarray}
where $\lambda_c$ is,
\begin{equation}
\lambda_c = \lim_{\tau\rightarrow \infty}
{\sum_{t=1}^{\tau} \ln|f^\prime (s_{t})|}
\end{equation}
The coupling strength range for which the d-synchronized solution
is stable, 1.e $\lambda_d < 0$, is given by,
\begin{equation}
1-\frac{1}{\exp{\lambda_c}} < \epsilon
\label{range-3-fx}
\end{equation}

{\it Coupling function $g(x)=f(x)$}:
We have investigated the network containing three nodes numerically
with $g(x)=f(x)$.
For logistic map with $\mu=4$, we find that nodes get d-synchronized
for coupling strength ranges
$0.15 < \epsilon < 0.2$ and $0.5 < \epsilon \le 1$. When we investigate further
these two $\epsilon$ ranges we find that in the lower range the
behaviour is mostly periodic while in the upper range it is periodic
in the middle portion and is chaotic at both ends. For $0.5
< \epsilon \le 1$ it is easy to see why the synchronized dynamics
gives a stable attractor. Since $\ln(1-\epsilon) <-\ln 2$ for
$\epsilon >0.5$, from Eq.~(\ref{lambda-3-fx}) we see that till
$\lambda_c$ is less than $\ln 2$, the value of Lyapunov exponent
for an isolated logistic map at $\mu=4$, the synchronized solution will be
stable. The stability of the d-synchronized state for $0.15 <
\epsilon <0.2$ appears to be because of either periodic attractor or chaotic
attractor close the periodic attractor with very small value of $\lambda_c$.

Now we discuss the special cases where nodes are synchronized with
variables showing periodic or fixed point behaviour. For this study
we use the original variables.

\begin{figure*}
\includegraphics[width=8cm]{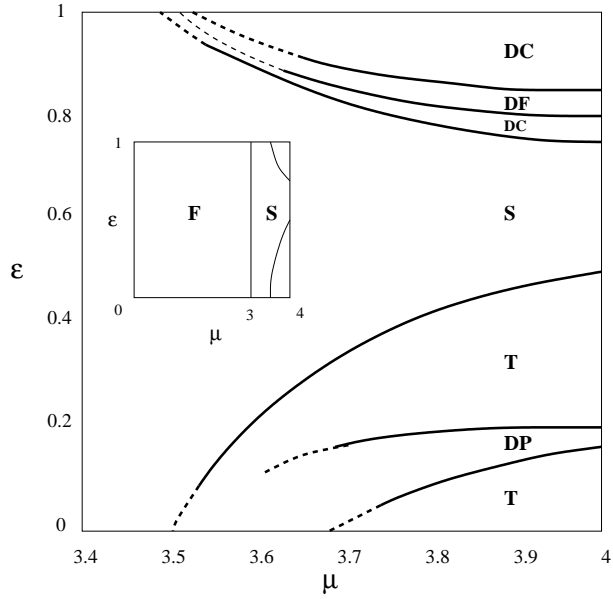}
\caption{ Phase space diagram showing different features of coupled dynamics 
in the two parameter space of $\mu$ and $\epsilon$ for three nodes
bipartite network with logistic map as local map
and coupling function $g(x)=f(x)$.  Different regions
are T. Turbulent region, DP. Driven periodic, DF. Driven fixed point,
DC. Driven Chaotic, S. Self organized region and F. Fixed point.
Region boundaries are determined 
based on the asymptotic behaviour using several initial conditions, 
synchronization behaviour and the largest Lyapunov exponent.
The dashed lines indicate uncertainties in 
determining the boundaries. The inset shows the phase diagram for the entire  
range of parameter $\mu$ from 0 to 4.}
\label{phase-3-fx}
\end{figure*}

{\it Case I. Synchronization to Fixed point}:
Nodes get synchronized to a fixed point
with one set of nodes having one value and the other set
of nodes having a different value such as $(x^1_t = x^2_t = X^{\star}_1)$
and ($x^3_t = X^{\star}_2$).
Eigenvalues of Jacobian matrix (\ref{jacobian-3}) with $g(x)=f(x)$ are,
\begin{subequations}
\begin{eqnarray}
\Lambda_1 &=& (1-\epsilon) f^{\prime}_1
\label{Lambda1-case1-3-fx} \\
\Lambda_{2,3} &=& \frac{1-\epsilon}{2}(f^{\prime}_1 + f^{\prime}_2)
 \nonumber \\
 & & \pm \frac{1}{2}\sqrt{(1-\epsilon)^2 (f^{\prime}_1+f^{\prime}_2)^2 - 
4(1-2\epsilon)f^{\prime}_1 f^{\prime}_2}
\label{Lambda-case1-3-fx}
\end{eqnarray}
\end{subequations}
where $f^\prime_1$ and $f^\prime_2$ are the derivatives at $X_1$ and $X_2$
respectively.
By putting the condition that the
magnitude of the above eigenvalues should be less than one,  we get
the $\epsilon$ range for which the fixed point state is stable.
When the first two nodes synchronize with each other, the three coupled maps
system behaves like just two coupled maps and within this $\epsilon$ range 
all the solutions are those of
the two coupled maps. 
The expressions  for $X^{\star}_1$ and $X^{\star}_2$ are
\begin{equation}
X^{\star}_{1,2} = \frac{ (1-\mu+2\mu \epsilon) \pm 
\sqrt{ (1-\mu+2\mu \epsilon)^2 - 4\epsilon(1 - \mu
+ 2 \mu \epsilon)}}{2 \mu (2\epsilon-1)}
\label{soln-case1-3-fx}
\end{equation}
From Eqs.~(\ref{Lambda1-case1-3-fx}) and~(\ref{Lambda-case1-3-fx}) the 
coupling strength values for which the fixed point solution is
stable must satisfy \[ |(1-\epsilon) f^\prime_1| < 1 \]
and the following condition
\begin{widetext}
\begin{equation}
\frac{1}{2}\left(1 + \sqrt{\frac{3}{\mu (\mu-2)}}\right)  <  \epsilon < 
\frac{3+2 \mu (\mu-2)+\sqrt{(3+2 \mu(\mu-2))^2-4 \mu(\mu-2)(\mu-1)^2}}
{4 \mu (\mu-2)};. \label{range1-3-fx}
\end{equation}
\end{widetext}

{\it Case II. Synchronization to period two orbit}:
Consider a periodic orbit 
of period two where the first two nodes are d-synchronized and
have the value $X_1^p$, and the third node has
has the value $X_2^p$. The period two orbit is obtained by
interchanging these two values at successive
time steps. The Jacobian matrix for this periodic orbit can be written as 
$J =J_1 J_2$ where
\begin{equation}
J_1 = \left( \begin{array}{cccc} 
(1-\epsilon) f^{\prime}_1 & 0 & \epsilon f^{\prime}_2  \\
0 & (1-\epsilon) f^{\prime}_1 & \epsilon f^{\prime}_2 \\
\frac{\epsilon}{2} f^{\prime}_1 & \frac{\epsilon}{2} f^{\prime}_1 
& (1-\epsilon) f^{\prime}_2
\end{array} \right)
\label{J1-3-fs-case2}
\end{equation}
where $f_1^\prime$ and $f_2^\prime$ are derivatives at $X_1^p$ and
$X_2^p$ respectively
and $J_2$ is obtained by interchanging suffixes 1 and 2 in the
expression for $J_1$.
Eigenvalues of $J$ are given by
\begin{subequations}
\begin{eqnarray}
\Lambda_1 &=& (1-\epsilon)^2 f^{\prime}_1 f^{\prime}_2
\label{Lambda1-3-fx} \\
\Lambda_{2,3} &=&  \left(\frac{\epsilon}{2}(f^{\prime}_1 + f^{\prime}_2) 
\pm \sqrt{\frac{\epsilon^2}{4} (f^{\prime}_1+f^{\prime}_2)^2 +
(1-2\epsilon)f^{\prime}_1 f^{\prime}_2 } 
\right)^2
\label{Lambda2-3-fx}
\end{eqnarray}
\end{subequations}
For $f(x)$ given by logistic map, the periodic points are
\begin{equation}
X^p_{1,2} = s \pm \sqrt{s(1-s-\frac{1}{\mu})},
\label{X12-case2-3-fx}
\end{equation}
where \[s = \frac{1}{2}\left( 1 + \frac{1}{\mu(1-2\epsilon)}\right).\]
After imposing the condition $\left | \Lambda_{1,2,3} \right | < 1$
we can get the $\epsilon$
range for which the period two orbit is stable. The first eigenvalue
gives us the condition that
\begin{equation}
1-\epsilon <
\frac{|1-2\epsilon|}{\sqrt{(1+\mu-2\epsilon\mu) 
(4\epsilon-1+\mu-2\epsilon\mu)}}.
\end{equation}
 The other two
eigenvalues give us the same condition as in Eq. (\ref{range1-3-fx})
except that $\epsilon$ in this inequality is substituted by
$1-\epsilon$ similar to the case of two coupled maps \cite{driven-T1S2}.
For logistic map with $\mu=4$, this range is given by 
$0.16.. < \epsilon < 0.20..$.

{\it Case III. Synchronization of all three nodes}:
It is also possible that all the three nodes get 
synchronized (s-synchronization). The eigenvalues 
of Jacobian matrix for this synchronized state
($x^1_t=x^2_t=x^3_t = x_t$) 
can simply be written from Eqs.~(\ref{Lambda1-case1-3-fx}) 
and~(\ref{Lambda-case1-3-fx}), by putting 
$f^\prime_1=f^\prime_2=f^{\prime}(x_t)$, which gives Lyapunov
exponents as,
\begin{subequations}
\begin{eqnarray}
\lambda_1 &=& \lambda_u = \frac{1}{\tau} \lim_{\tau to \infty} 
\sum_{t=1}^\tau \ln|f^{\prime}(x_t)|\\
\lambda_2 &=& \ln(1-\epsilon) + \lambda_u\\
\lambda_3 &=& \ln(1-2 \epsilon) + \lambda_u
\label{lya-self-3-fx}
\end{eqnarray}
\end{subequations}
where $\lambda_u$ is the Lyapunov exponent for an uncoupled map $f(x)$.
For synchronous orbits to be stable, the two Lyapunov exponents
($\lambda_2$ and $\lambda_3$) 
corresponding to the transverse eigenvectors should be negative. We get
the following $\epsilon$ range for which both the transverse eigenvalues are negative,
\begin{equation}
1- e^{-\lambda_u} < \epsilon < \frac{1}{2}(1 + e^{-\lambda_u})
\label{range3-3-fx}
\end{equation}
In this region coupled dynamics may lie on a chaotic or a periodic attractor
depending upon the value of $\lambda_u$.
For $f(x)=\mu x(1-x)$ with $\mu=4$, $\lambda_u=\ln 2$ and thus the
range of stability of the s-synchronization of all three
nodes is (0.5, 0.75) and the dynamics is chaotic.

\begin{figure}
\centerline{
\includegraphics[width=11cm]{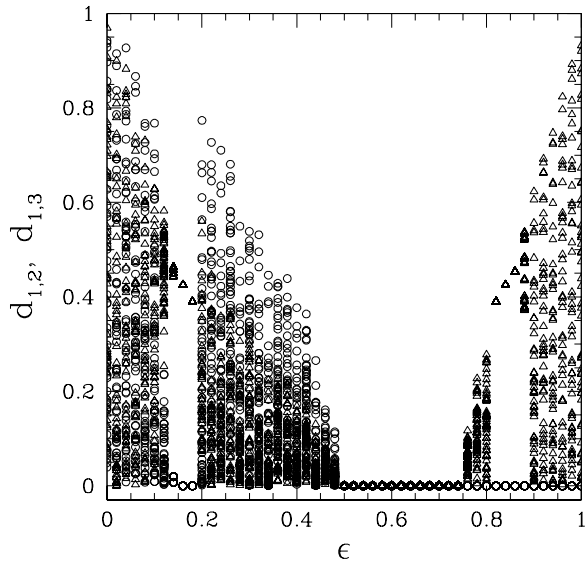}}
\caption{The figure shows the variation of
two sets of difference, $|x^1_t-x^2_t|$ (open circles)
and $|x^1_t-x^3_t|$ (crosses) for a three nodes network as a function of the 
coupling strength
$\epsilon$ for $f(x)=\mu x (1-x)$ with $\mu=4$ and $g(x)=f(x)$. 
For each $\epsilon$,
100 values of the differences are plotted after an initial transient.}
\label{diff-3-fx}
\end{figure}

\begin{figure*}
\centerline{\begin{tabular}{cc}
\includegraphics[width=11cm]{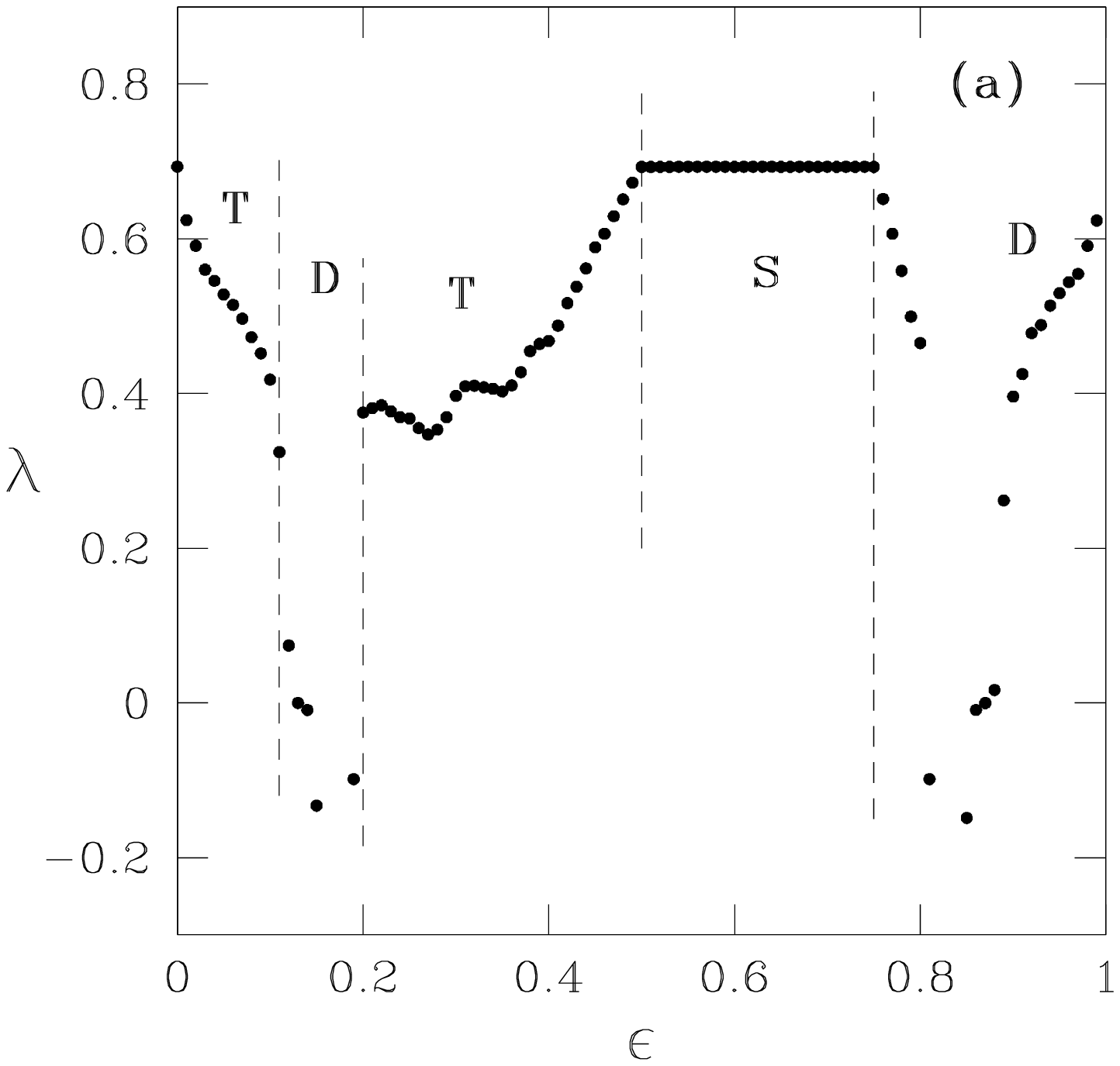} & \includegraphics[width=11cm] {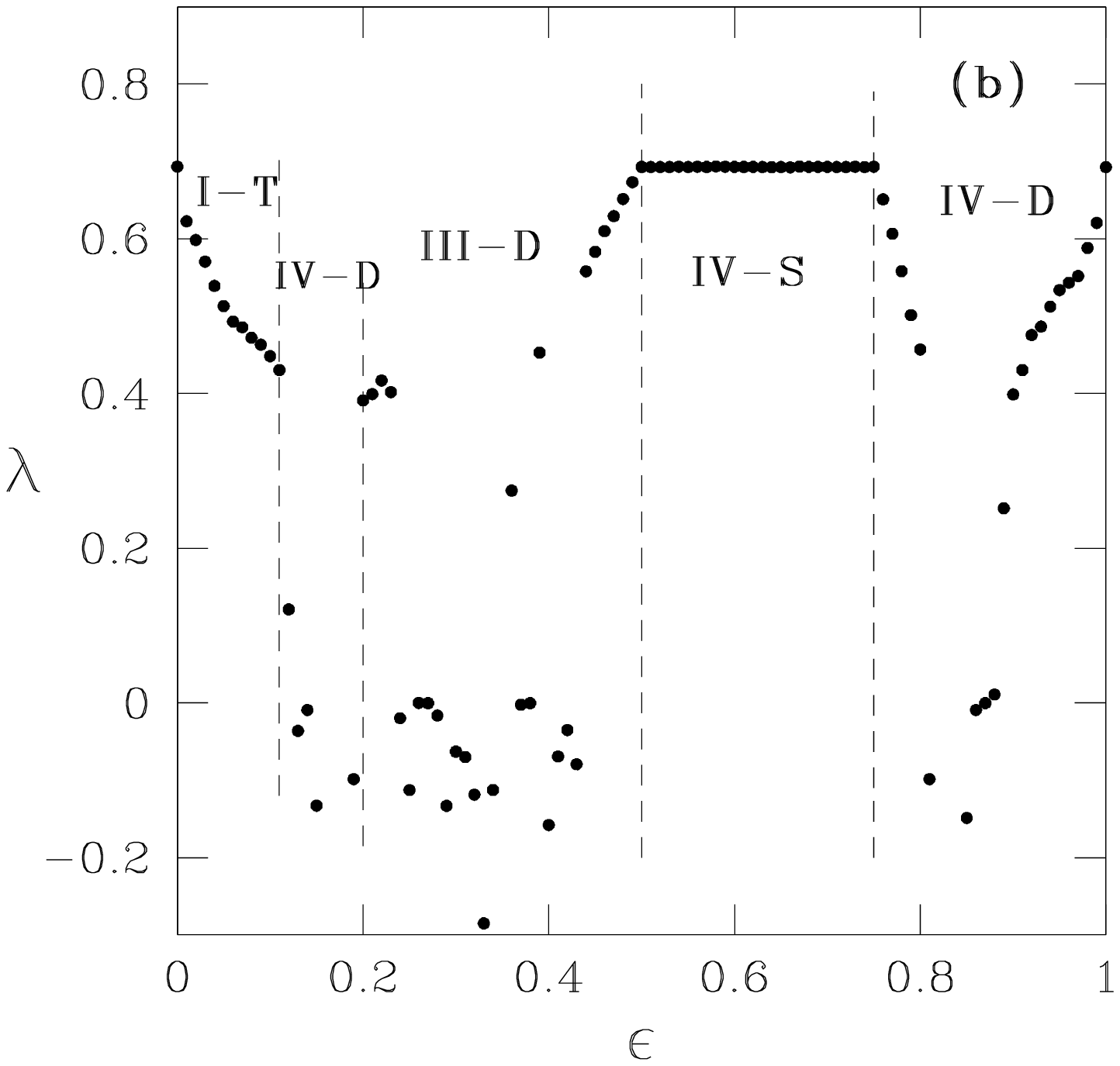}
\end{tabular} }
\caption{
(a) The figure shows the largest
Lyapunov exponent $\lambda$ as a function of the coupling strength $\epsilon$
for the three nodes bipartite network with logistic map as
local map with $\mu=4$ and $g(x)=f(x)$. (b) Same as for 
(a) but for a bipartite network of 50 nodes with $N1=N2=25$.}
\label{lya-3-fx}
\end{figure*}

\begin{figure*}
\centerline{\begin{tabular}{cc}
\includegraphics[width=11cm]{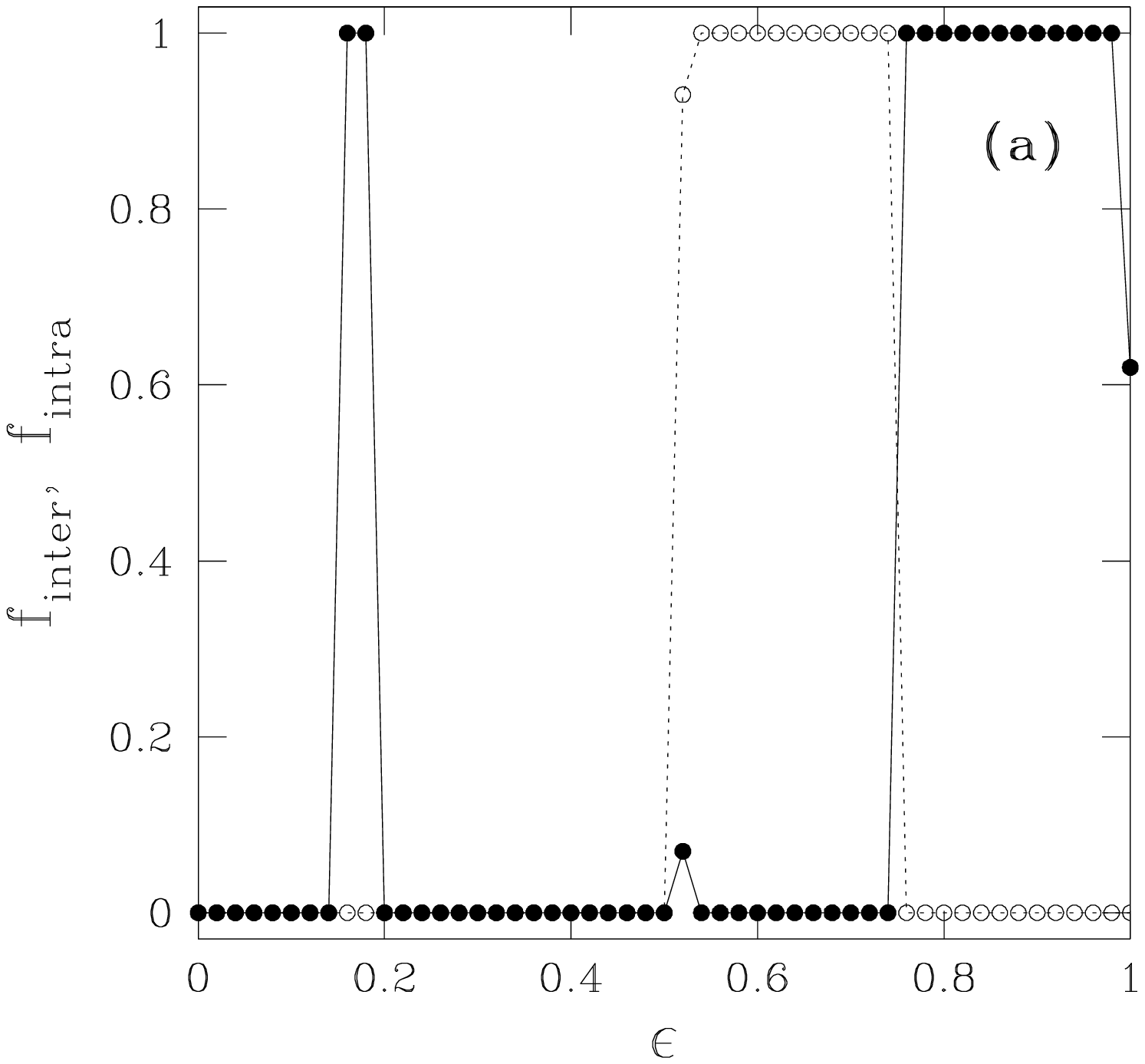} & \includegraphics[width=11cm]{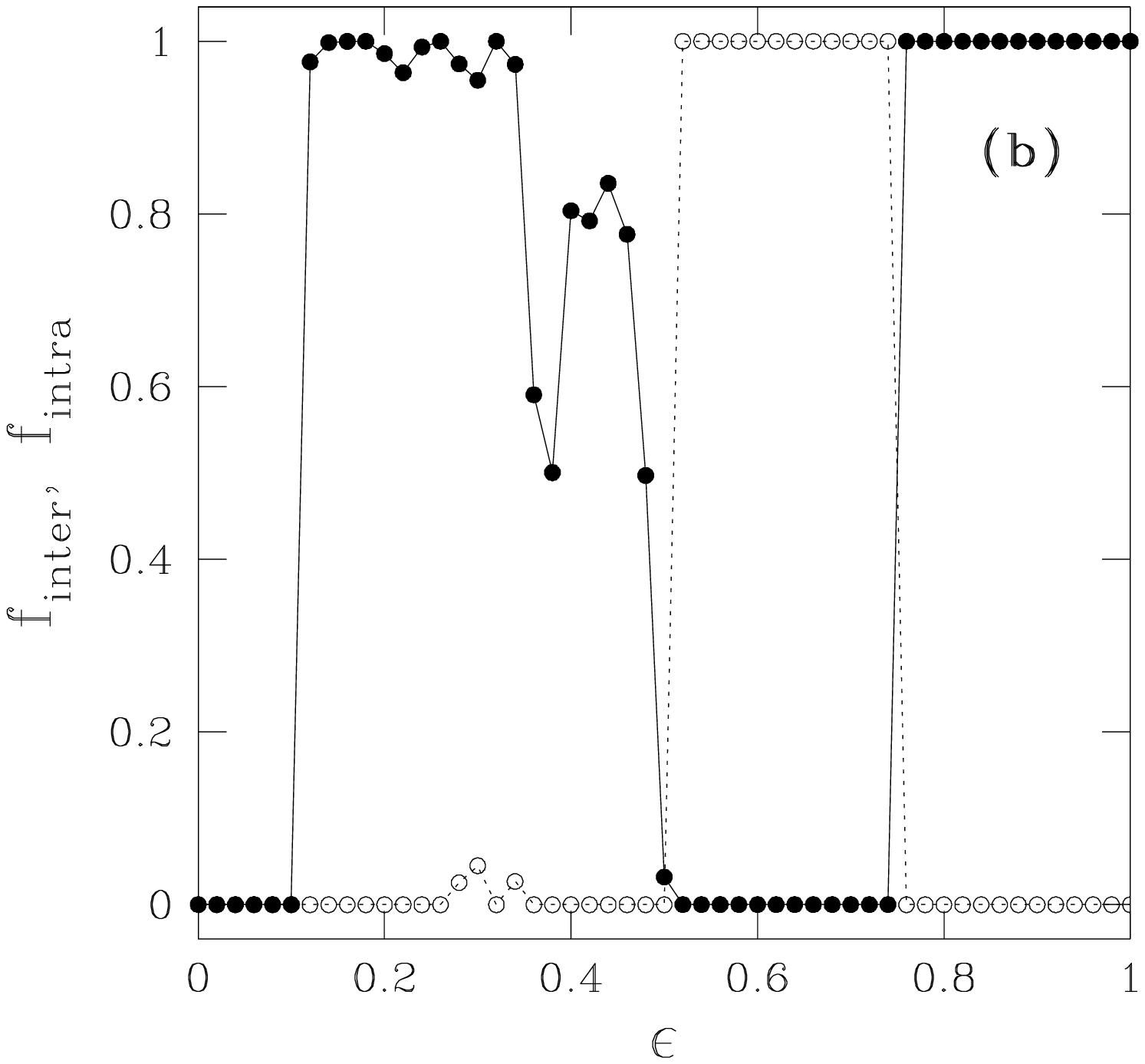}
\end{tabular} }
\caption{
(a) The figure shows the fractions of inter- and intra-cluster couplings,
$f_{inter}$ and $f_{intra}$, 
as a function of the coupling strength $\epsilon$
for the three nodes bipartite network, logistic map as
local map with $\mu=4$ and $g(x)=f(x)$. The values are obtained by
averaging over 50 random initial conditions. (b) Same as for 
(a) but for a bipartite network of 50 nodes with $N1=N2=25$.}
\label{inter-intra-3-fx}
\end{figure*}

{\it Phase diagram in $\mu-\epsilon$ plane}:
Fig.~\ref{phase-3-fx} shows different phases in the $\mu-\epsilon$
plane for three nodes bipartite network with $g(x)=f(x)$. For $\mu<3$
we get a fixed point solution. To understand
the remaining phase diagram consider the line $\mu=4$.
Fig.~\ref{diff-3-fx} shows two sets of differences between
the values
of variables, $|x^1-x^2|$ (open circles) and $|x^1-x^3|$ (crosses) as
a function of the coupling strength $\epsilon$. Bipartite d-synchronized
state and global s-synchronized state are clearly seen. 
Fig.~\ref{lya-3-fx}(a) 
shows largest Lyapunov exponent and Fig.~\ref{inter-intra-3-fx}(a) shows the fractions
of inter- and intra- couplings, $f_{inter}$
and $f_{intra}$, as a function of $\epsilon$ for $mu=4$.
The two quantities $f_{\rm intra}$ and $f_{\rm inter}$ act as
measures for the intra-cluster couplings and the inter-cluster couplings
and are defined as \cite{sarika-REA1,sarika-REA2},
\begin{subequations}
\begin{eqnarray}
f_{\rm intra} &=& \frac{N_{\rm intra}}{N_c}\\
f_{\rm inter} &=& \frac{N_{\rm inter}}{N_c}
\end{eqnarray}
\end{subequations}
where $N_{\rm intra}$ and $N_{\rm inter}$ are the numbers of intra- and
inter-cluster couplings respectively. In $N_{\rm inter}$, couplings
between two isolated nodes are not included.
Initially for small coupling strength values nodes are in turbulent region with
no cluster formation at all (region T). As the coupling strength increases
beyond a critical $\epsilon_c$
we get bipartite d-synchronized state (region DP). 
This region corresponds to the Case II of period two orbits discussed above
in this subsection.
When the coupling strength increases further we get a reappearance of
turbulent region. 
As the coupling strength is increased further all the nodes are synchronized
giving one cluster with global s-synchronization (region S). 
This region corresponds to the case III discussed above and the range of
$\epsilon$ for this region is given by
Eq.~(\ref{range3-3-fx}). In this region the coupled dynamics lies on
a chaotic attractor. In the last three regions (DC, DP and DC), we get 
driven bipartite synchronization.
The middle region (DP) displays driven fixed point solution (case I) while
the other regions show chaotic behaviour.

For $\mu < 4$, the coupling  strength range where
all the nodes are synchronized (region S) increases in size and the coupling
strength ranges where nodes show d-synchronization
shift towards the nearest boundaries $\epsilon=0$ and $\epsilon=1$.

{\it Coupling function $g(x) = x$}:
Numerical analysis shows periodic solutions which we now consider
in detail.
When the first two nodes are d-synchronized the solutions
of the dynamical equations are similar to the two nodes case.
This simplifies the analysis considerably.

\begin{figure}
\centerline{
\includegraphics[width=8cm]{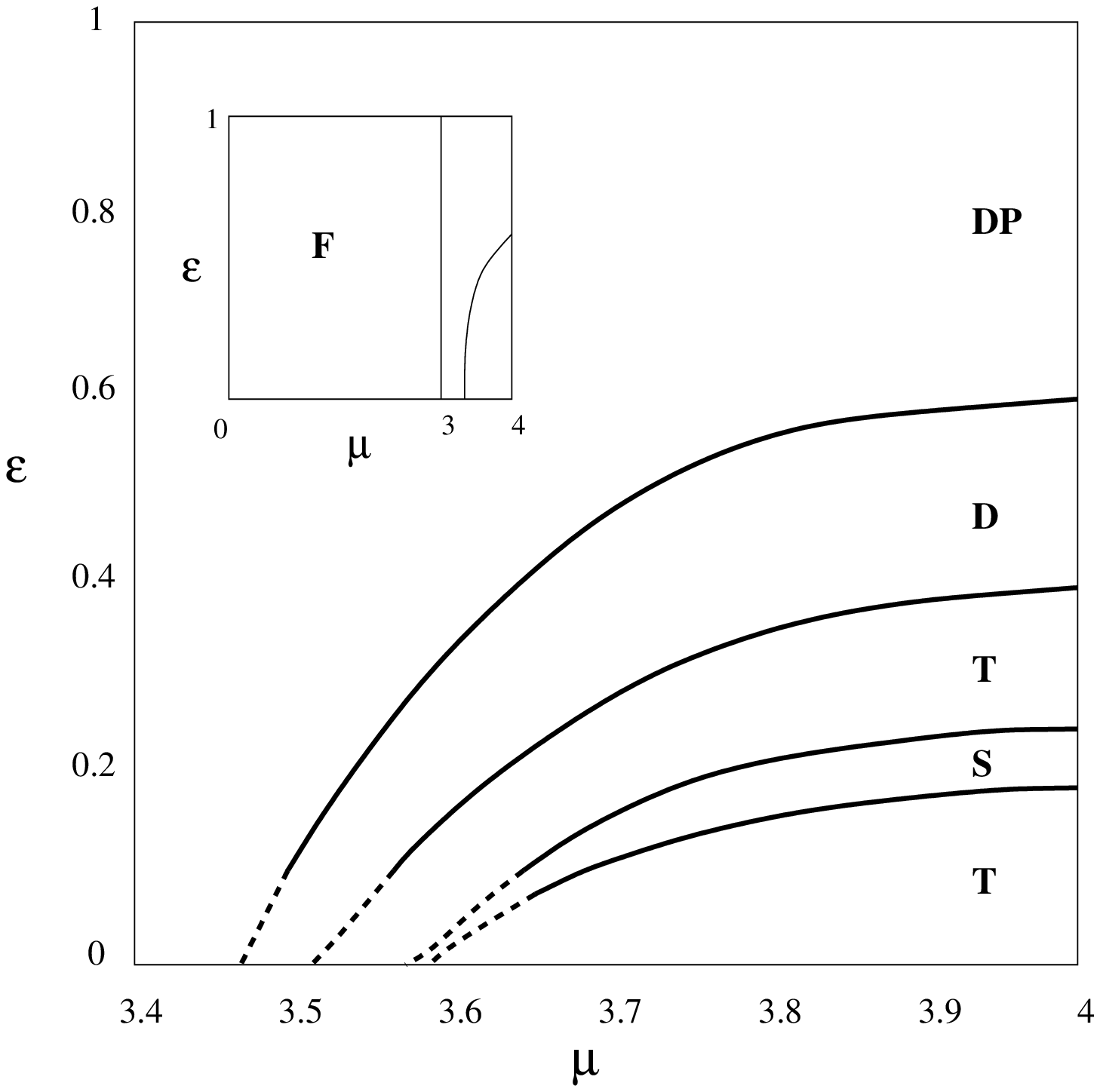}}
\caption{
Phase space diagram showing different features of coupled dynamics in
the two parameter space of $\mu$ and $\epsilon$ for three nodes
bipartite network with logistic map as local map
and coupling function $g(x)=x$.
Different regions
are T. Turbulent region, DP. Driven periodic, DF. Driven fixed point,
DQ. Driven Quasiperiodic, DC. Driven Chaotic, S. Self organized region and F. Fixed point.
Region boundaries are determined 
based on the asymptotic behaviour using several initial conditions, 
synchronization behaviour and the largest Lyapunov exponent.
The dashed lines indicate uncertainties in 
determining the boundaries. The inset shows the phase diagram for the entire  
range of parameter $\mu$ from 0 to 4.}
\label{phase-3-x}
\end{figure}

{\it Case I. Driven synchronization to period two orbit}:
First we consider the case where the coupled dynamics shows periodic
attractor with period two behaviour 
and the variables take the values
\begin{eqnarray}
x^1_t = x^2_t  = x^3_{t+1} & = & X_1^p  \nonumber \\
x^3_t = x^1_{t+1} = x^2_{t+1} &=& X^p_2 
\label{cond-period-3-x}
\end{eqnarray}
Using the product of Jacobians of Eq.~(\ref{jacobian-3}) for two consecutive
time steps the eigenvalues for this periodic orbit can be easily obtained. 
The eigenvalue $\Lambda_d$ associated with
the difference variable $d_t$ is given by
\begin{equation}
\Lambda_d = (1-\epsilon)^2f_1^{\prime} f^{\prime}_2,
\end{equation}
The other two eigenvalues are the eigenvalues of product matrix $J_1 J_2$,
where $J_1$ is given by Eq.~(\ref{jacobian-case2-2-x}) and hence the 
eigenvalues are given by Eq.~(\ref{Lambda-case2-2-x}).
The solution of the periodic orbit is the same as for two coupled maps 
(Eq.~(\ref{sol-case2-2-x})). Using all the three
eigenvalues of Jacobian matrix, the coupling strength range,
for which the periodic orbit given by Eq.~(\ref{cond-period-3-x})
is stable, is given by
\begin{equation}
\max \left(1 - \frac{1}{\sqrt{f^\prime_1 f^\prime_2}}, \frac{|f_1^{\prime} f_2^\prime| - 1}
{|f_1^\prime f_2^\prime| + 1}\right) <  \epsilon < 1
\label{range-case1-3-x}
\end{equation}
The periodic points $X_1$ and $X_2$ are also the periodic points of uncoupled
map $f$ (Eq.~(\ref{mu-case1-2-x})). For logistic map, we get the following range of coupling strength in terms of logistic map
parameter $\mu$,
\begin{equation}
1 - \frac{2}{\mu^2-2\mu-3}
< \epsilon <1
\label{range-logistic-case1-3-x}
\end{equation}
which is the same as Eq.~(\ref{range-case2-2-x}).
The lower bound of $\epsilon$ matches exactly with the boundary
between the 
regions IV-DP and IV-DQ of Fig.~1 of Ref.~\cite{sarika-REA2} and regions DP and DQ
of Fig.~\ref{phase-3-x}.

\begin{figure}
\centerline{
\includegraphics[width=11cm]{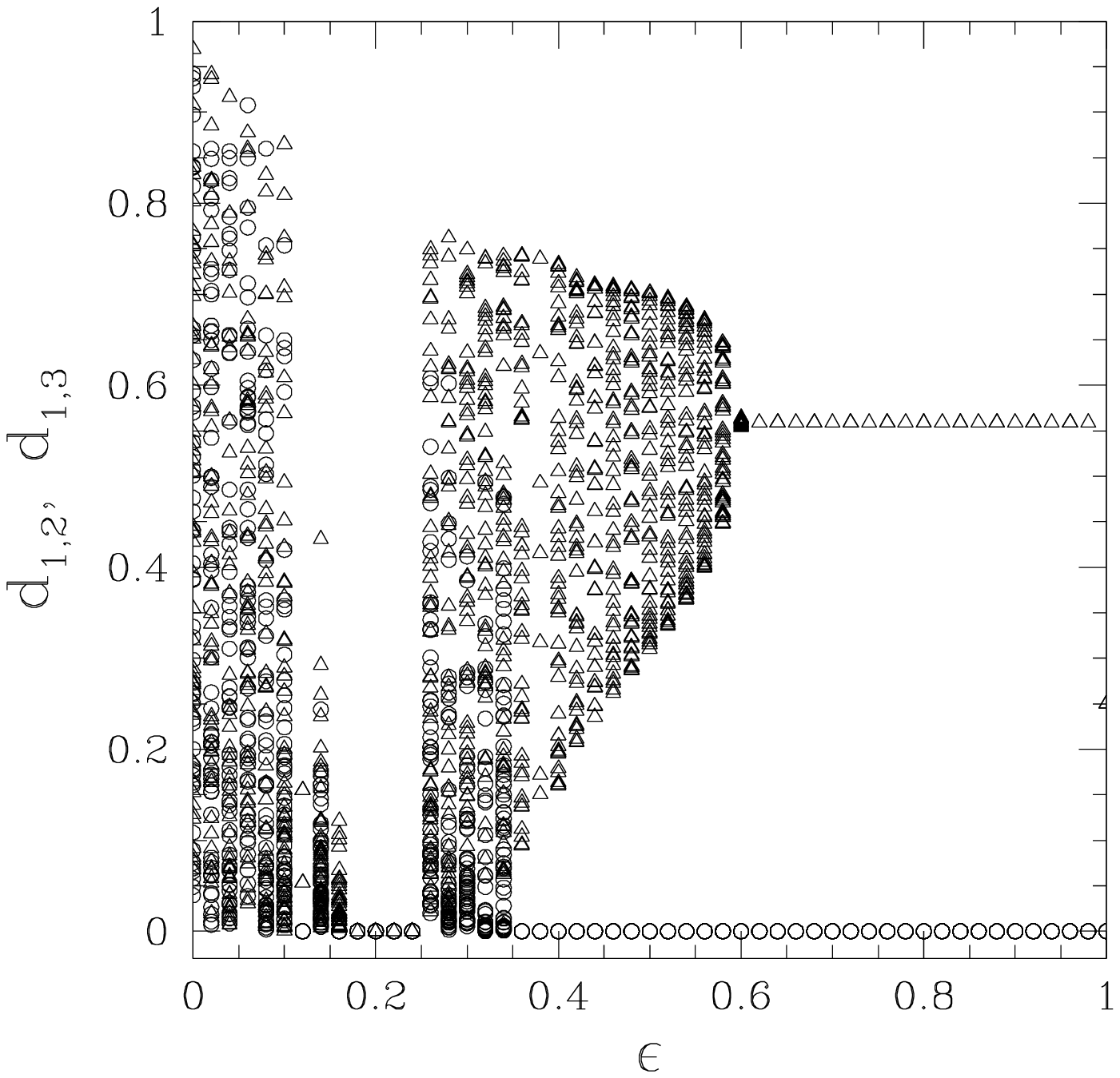}}
\caption{
The figure shows the variation of
two sets of difference, $|x^1_t-x^2_t|$ (open circles)
and $|x^1_t-x^3_t|$ (crosses) for a three nodes network as a function of the coupling strength
$\epsilon$ for $f(x)=\mu x (1-x)$ with $\mu=4$ and $g(x)=x$. 
For each $\epsilon$,
100 values of the differences are plotted after an initial transient.
}
\label{diff-3-x}
\end{figure}

{\it Case II. Synchronization of all three nodes}:
All three nodes get synchronized for a small coupling strength
region with dynamics lying on periodic orbits of period two such
that
\begin{eqnarray}
x^1_t = x^2_t = x^3_t = X^p_1 \nonumber \\
x^1_{t+1} = x^2_{t+1} = x^3_{t+1} = X^p_2
\label{cond-case2-3-x}
\end{eqnarray}
The eigenvalue of the Jacobian for this periodic orbit for the
difference variable is simply 
$(1-\epsilon)^2 f^\prime_1 f^\prime_2$ and the other two eigenvalues
are the same as for two coupled maps and are given by 
Eq.~(\ref{Lambda1-2-x}). The periodic points are given by
Eq.~(\ref{sol-case1-2-x}). The coupling strength
range for which all three nodes are synchronized
for logistic map with $\mu=4$, 
is $0.18...< \epsilon < 0.24...$ which is same as Case I for two
coupled maps with $g(x)=x$.  

\begin{figure*}
\centerline{\begin{tabular}{cc}
\includegraphics[width=11cm]{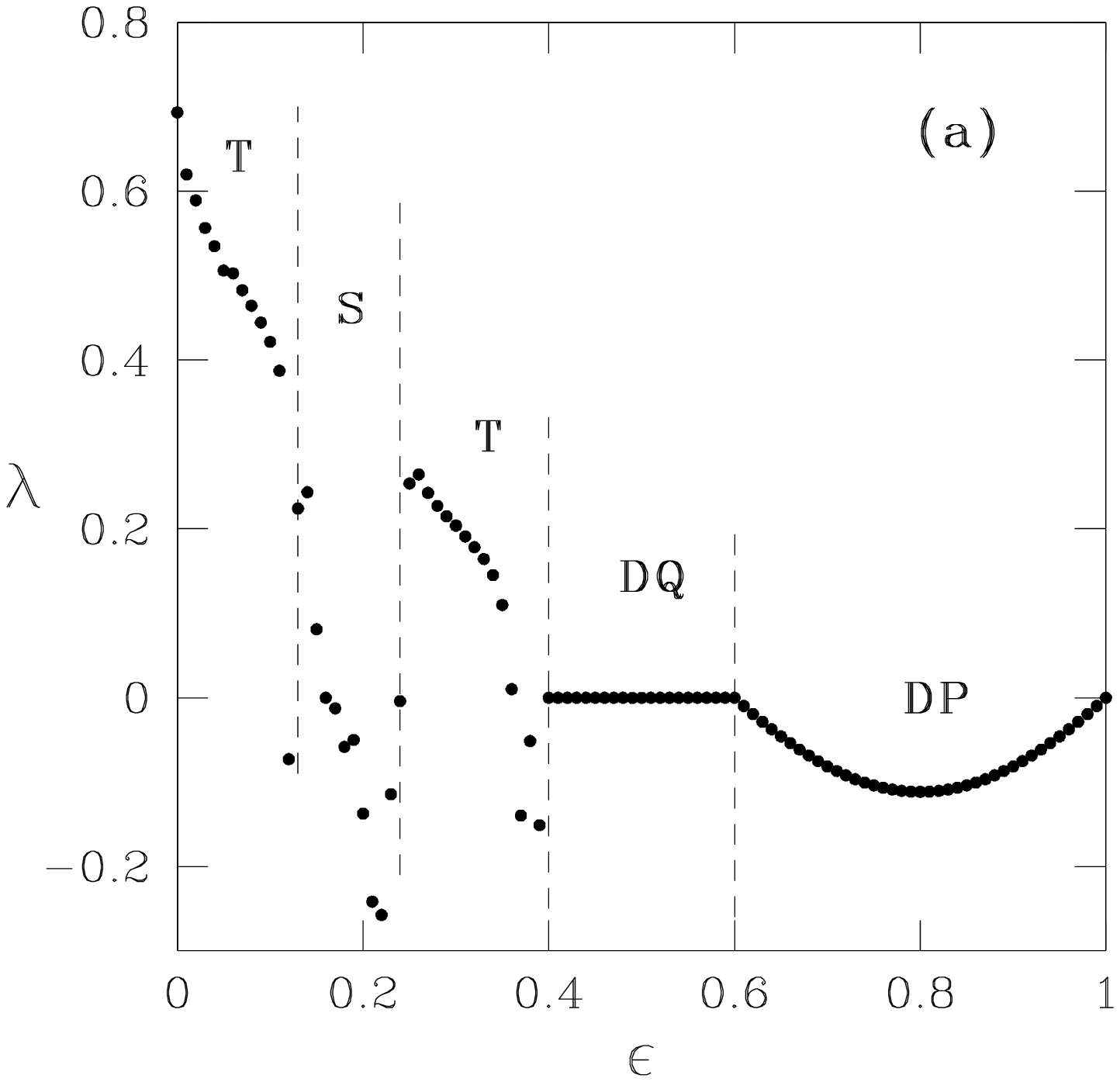} & \includegraphics[width=11cm]{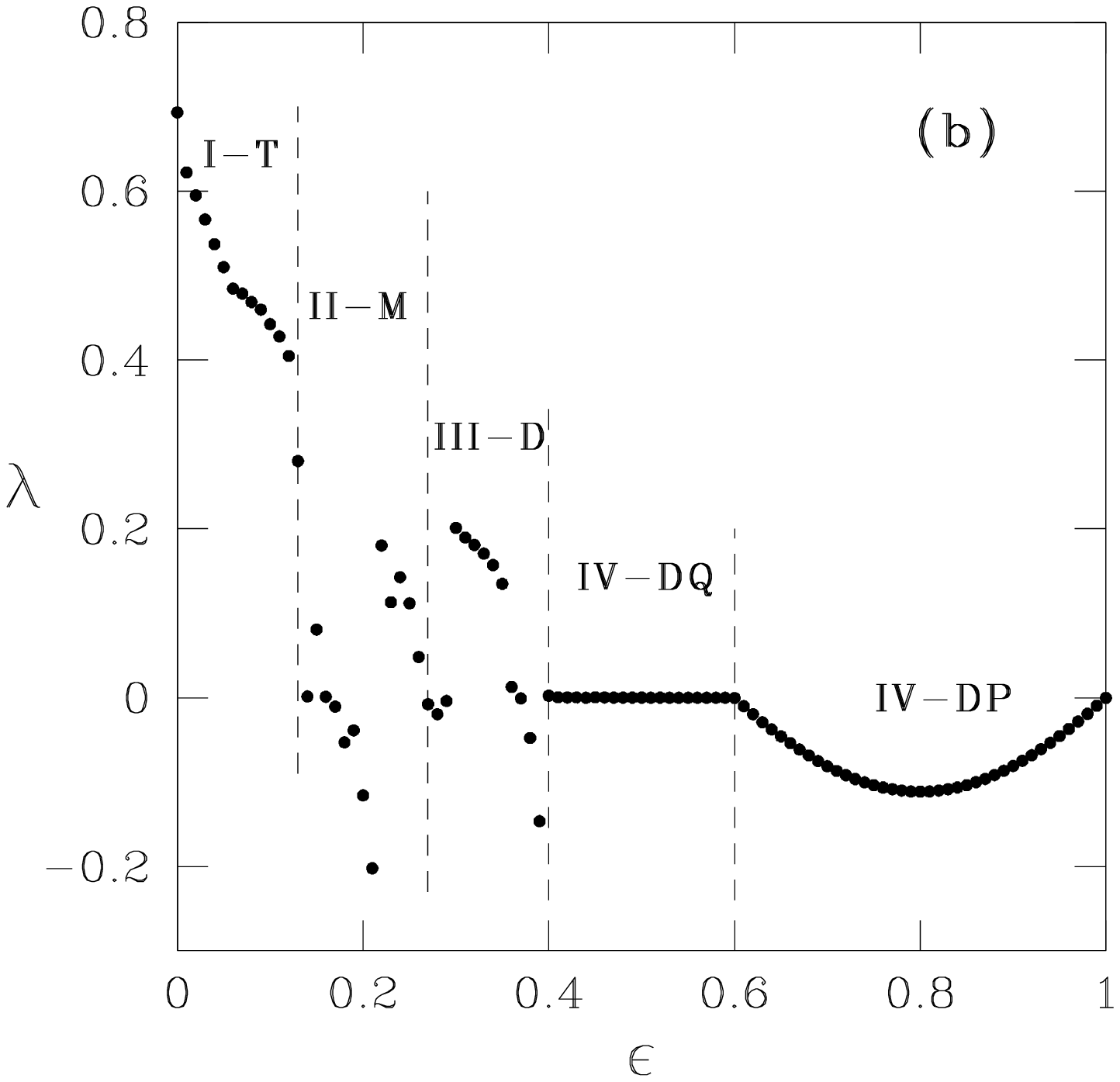}
\end{tabular} }
\caption{
(a) The figure shows the largest
Lyapunov exponent $\lambda$ as a function of the coupling strength $\epsilon$
for the three nodes bipartite network with logistic map as
local map with $\mu=4$ and $g(x)=x$. (b) Same as for 
(a) but for a bipartite network of 50 nodes with $N1=N2=25$.}
\label{lya-3-x}
\end{figure*}

{\it Phase diagram in $\mu- \epsilon$ space}:
Fig.~\ref{phase-3-x} shows different phases in the $\mu-\epsilon$
plane for three nodes bipartite network with $g(x)=x$. For $\mu<3$
we get a fixed point solution. To understand
the remaining phase diagram consider the line $\mu=4$.
Fig.~\ref{diff-3-x} shows two sets of differences between
the values
of variables, $|x^1-x^2|$ (open circles) and $|x^1-x^3|$ (crosses) as
a function of the coupling strength $\epsilon$. Bipartite d-synchronized
state and global s-synchronized state are clearly seen. 
Fig.~\ref{lya-3-x}(a) 
shows largest Lyapunov exponent and Fig.~\ref{inter-intra-3-x}(a) shows the fractions
of inter- and intra- couplings, $f_{inter}$
and $f_{intra}$, as a function of $\epsilon$ for $mu=4$.
Initially for small coupling strength values nodes are in turbulent region with
no cluster formation at all (region T). As the coupling strength increases
beyond a critical $\epsilon_c$,
we get global s-state (region S). 
This region is the case II considered above.
When the coupling strength increases further we get a reappearance of
turbulent region.
In the last two regions we get driven bipartite synchronization 
(regions DQ and DP).
The last region corresponds to case I of period two discussed above
in this subsection 
and the critical coupling strength for it
is given by  Eq.~(\ref{range-case1-3-x}).

For $\mu < 4$, the coupling strength region for
d-synchronization gets wider and that for s-synchronization gets thiner with a shift towards $\epsilon=0$.  

\subsection{Complete bipartite networks}
Let us now consider a complete bipartite network of
$N=N1+N2$ nodes and dynamics defined by Eq.~(\ref{model-driven}).
We define a bipartite synchronized state of the bipartite network as the one that 
has that all $N1$ elements of the first set synchronized
to some value, say $X_1(t)$, and 
all $N2$ elements of the second set  synchronized
to some other value, say $X_2(t)$.
Linear stability analysis of the bipartite synchronized state
can be done using the Jacobian matrix,
\begin{widetext}
\begin{equation}
J_t = \left ( \begin{array}{cccccccc}
(1-\epsilon) f^{\prime}_1 & 0& \dots & 0 & 
\frac{\epsilon}{N2} g^{\prime}_2& 
\frac{\epsilon}{N2} g^{\prime}_2& \dots & \frac{\epsilon}{N2} g^{\prime}_2 
\\
\vdots & \vdots & \vdots & \vdots & \vdots & \vdots & \vdots & \vdots\\
0 & 0 & \dots & (1-\epsilon) f^{\prime}_1 & \frac{\epsilon}{N2} g^{\prime}_2&
\frac{\epsilon}{N2} g^{\prime}_2 & \dots &
\frac{\epsilon}{N2} g^{\prime}_2
\\
\frac{\epsilon}{N1} g^{\prime}_1 & \frac{\epsilon}{N1} g^{\prime}_1 & 
\dots & \frac{\epsilon}{N1} g^{\prime}_1 &
(1-\epsilon) f^{\prime}_2 & 0 &  \dots & 0
\\
\vdots & \vdots & \vdots & \vdots & \vdots & \vdots & \vdots & \vdots\\
\frac{\epsilon}{N1} g^{\prime}_1 & \frac{\epsilon}{N1} g^{\prime}_1 &
 \dots & \frac{\epsilon}{N1} g^{\prime}_1 &
0 & 0 & \dots & (1-\epsilon) f^{\prime}_2  \\
\end{array} \right )
\label{Jt-driven}
\end{equation}
\end{widetext}
where $g^{\prime}_1$ and $g^{\prime}_2$ are 
the derivative of $g(x)$ at $X_1$ and $X_2$ respectively.
It is easy to see that the eigenvectors and eigenvalues of the above Jacobian
matrix can be divided into three sets, $A, B, C$, as, 
\begin{widetext}
\begin{center}
\begin{tabular}{cccccccc}
set & Eigenvectors & & Eigenvalue & & No of eigenvalues & &
condition\\
$A$ & $(\alpha_1 \dots \alpha_{N1}, 0 \cdots 0)$ & & $(1-\epsilon)f_1^{\prime}$ & &
$N1 - 1$ & & $ \Sigma{\alpha}=0$\\
$B$ & $(0 \dots 0,\beta_1 \dots \beta_{N2})$ & & $(1-\epsilon)f_2^{\prime}$ & &
$N2 - 1$ & & $\Sigma{\beta} = 0$ \\
$C$ & $(\alpha, \dots \alpha, \beta, \dots \beta)$ & & - & & 2 && - 
\end{tabular}
\end{center}
\end{widetext}
Here, ($\alpha_1, \ldots \alpha_{N1}$), ($\beta_1, \ldots \beta_{N2})$
and $\alpha, \beta$ are complex numbers satisfying the conditions
specified in the last column.
The two eigenvalues corresponding to the set $C$ are the
eigenvalues of
the matrix
\begin{equation}
\left( \begin{array}{ccc}
(1-\epsilon)f_1^\prime & \epsilon g_2^\prime \\ 
\epsilon g_1^\prime &  (1-\epsilon)f_2^\prime 
\end{array} \right)
\end{equation}

We make the following observations.
(a) The three sets of eigenvectors, $A, B, C$, are
orthogonal to each other and the total space of eigenvectors may be
written as a direct sum of these three sets. (b) The three sets
of eigenvectors do not mix with each other under time evolution.
(c) The eigenvectors belonging to the first two sets ($A$ and $B$) 
are also the eigenvectors of the product of any number of Jacobian
matrices under time evolution. (d) The two sets of eigenvectors $A$ and $B$ 
are transverse to the synchronization manifold which is defined by the
last set of eigenvectors $C$.

Lyapunov exponents corresponding to the transverse eigenvectors
can be easily written as
\begin{subequations}
\begin{eqnarray}
\lambda_i &=& \ln|(1-\epsilon)| + \frac{1}{\tau} \lim_{\tau \to\infty}
\sum^\tau_{t=1} \ln |f^{\prime}(X_1)|, \nonumber \\
    & & \, \, \, \, \, {\rm for} \, i=1,\ldots,N1-1 \\
\lambda_i &=& \ln|(1-\epsilon)| + \frac{1}{\tau} \lim_{\tau \to\infty}
\sum^\tau_{t=1} \ln |f^{\prime}(X_2)|, \nonumber \\
   & & \, \, \, \, \, {\rm for} \, i=N1,\ldots,N-2.
\end{eqnarray}
\label{lya-driven-Nlarge}
\end{subequations}
The synchronized state is stable provided the transverse Lyapunov
exponents are negative.
If $f^\prime$ is bounded as is the case for the logistic map then from
Eqs.~(\ref{lya-driven-Nlarge}) we see that for $\epsilon$ larger than
some critical value, $\epsilon_c (<1)$, bipartite synchronized state will be
stable. Note that this bipartite synchronized state will be stable even if one
or both the remaining Lyapunov exponents corresponding to the set $C$ are
positive, i.e. the trajectories are chaotic.

Now we study the periodic behaviour of coupled dynamics of
Eq.~(\ref{model-driven}). Here we consider two cases, fixed point
attractor and periodic attractor with period two, as we studied for
three nodes.

\subsubsection{Coupling function $g(x)=f(x)$}
The fixed point bipartite synchronized solution with one set of nodes taking
value $X_1^*$ and the other set $X_2^*$, is given by
Eq.~(\ref{soln-case1-3-fx}). Jacobian matrix for this solution gives
three sets of eigenvalues, first set of $N1-1$ degenerate eigenvalues
$(1-\epsilon)f^\prime_1$, second set of $N2-1$ degenerate eigenvalues
$(1-\epsilon)f^\prime_2$ and third set of two eigenvalues given by
Eq.~(\ref{Lambda-case1-3-fx}). The conditions for the stability of
this solution is given by Eq.~(\ref{range1-3-fx}) and $|(1-\epsilon)
f^\prime_1,2| < 1$.

The bipartite synchronous period two solution is obtained when one set
of nodes take the value $X^p_1$ and the other set of nodes take the
value $X^p_2$ and the two values alternate in time. This solution is
the same as given by
Eq.~(\ref{X12-case2-3-fx}). The eigenvalues (Eqs.~(\ref{Lambda1-3-fx})
and~(\ref{Lambda2-3-fx}), with the $N-2$ fold
degeneracy of $\Lambda_1$)
and the conditions for stability of the solution are the same as for
Case II of three nodes with $g(x)=f(x)$.

The global synchronized solution where all nodes of the bipartite
network are synchronized, has the same Lyapunov exponents (except the
degeneracy) as in Eqs.~(\ref{lya-self-3-fx}) and the stability
criterion is again given by Eq.~(\ref{range3-3-fx}).     

\subsubsection{Coupling function $g(x)=x$}
Consider a periodic orbit of period two with bipartite synchronized
state where all nodes of one set take value
$X_1$ and and the other set $X_2$ and two values alternate in time. 
The solutions is same as for case I of three nodes bipartite network.
The stability ranges are given by Eqs.~(\ref{range-case1-3-x}) 
and~(\ref{range-logistic-case1-3-x}). 

\begin{figure*}
\centerline{\begin{tabular}{cc}
\includegraphics[width=11cm]{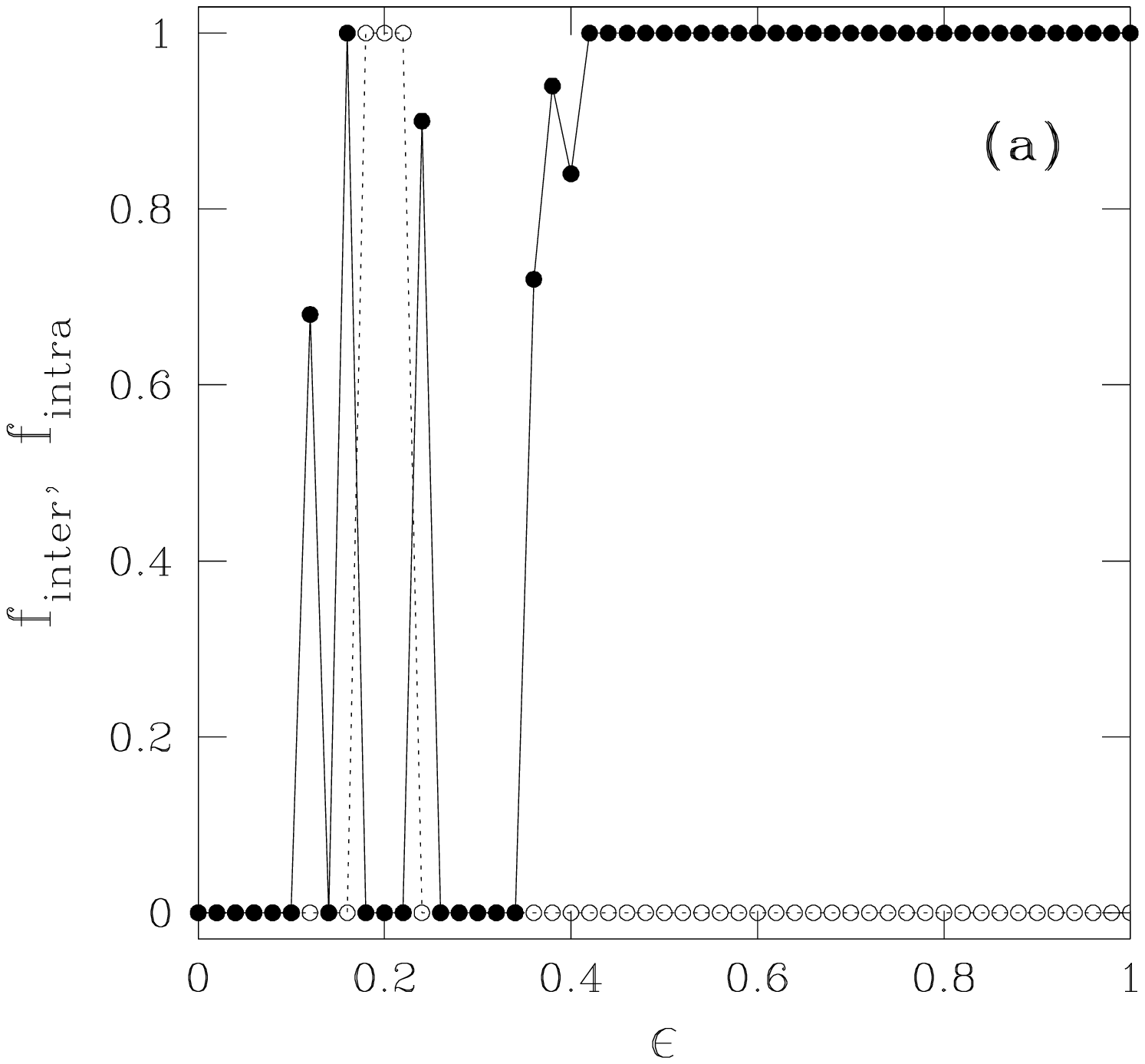} & \includegraphics[width=11cm]{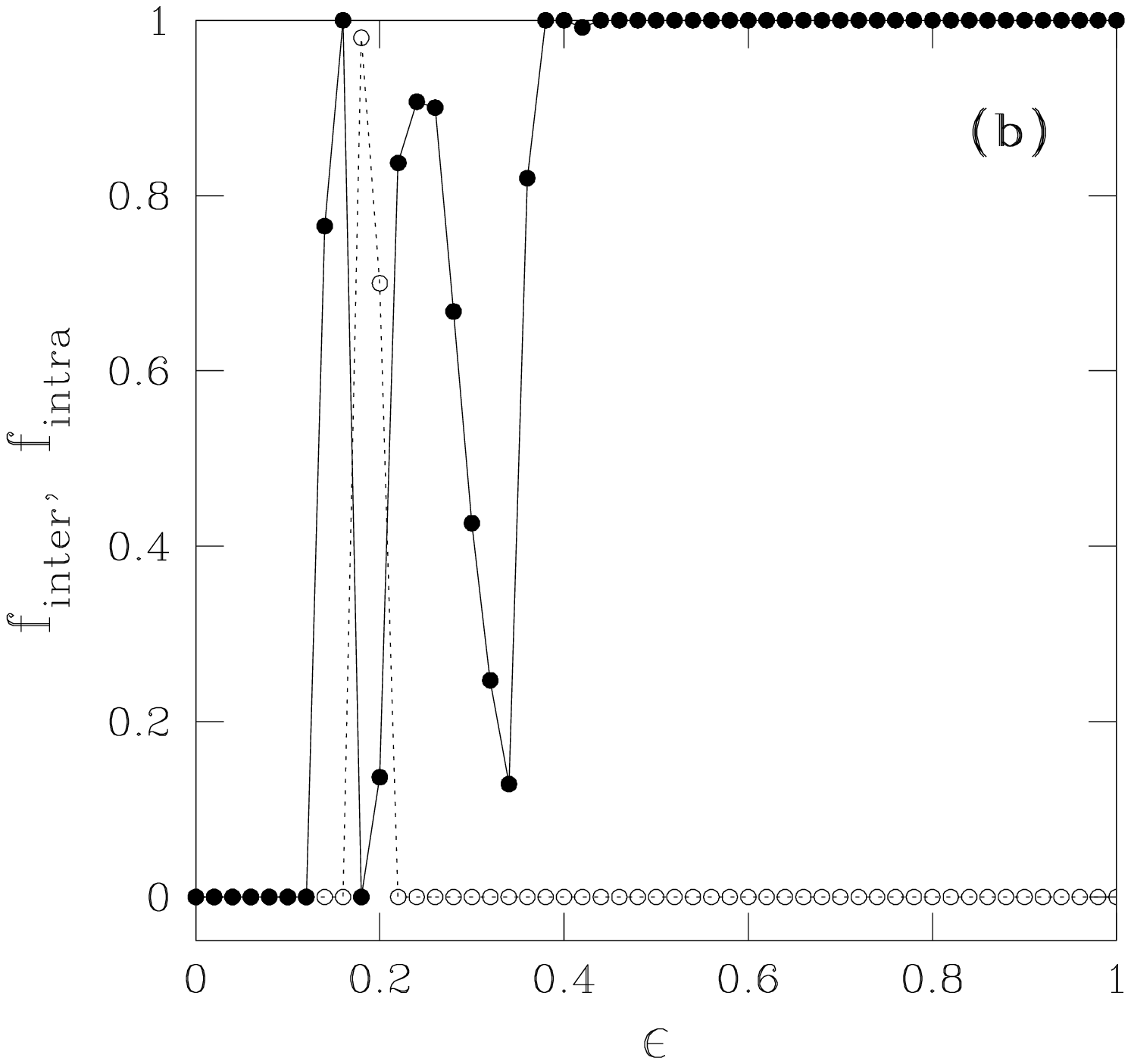}
\end{tabular} }
\caption{
(a) The figure shows the fractions of inter- and intra-cluster couplings,
$f_{inter}$ and $f_{intra}$, 
as a function of the coupling strength $\epsilon$
for the three nodes bipartite network, logistic map as
local map with $\mu=4$ and $g(x)=x$. The values are obtained by
averaging over 50 random initial conditions. (b) Same as for 
(a) but for a bipartite network of 50 nodes with $N1=N2=25$.}
\label{inter-intra-3-x}
\end{figure*}

\subsection{Lyapunov Function Analysis}
For the complete bipartite networks, Lyapunov function
as defined by Eq.~(\ref{Lya-fun}), for any two nodes belonging to the same
set is given by
\begin{equation}
V^{ij}_{t+1} = [ (1-\epsilon) (f(x^i_t) - f(x^j_t)) ]^2   
\label{Lyafun-3}
\end{equation}
Expanding around $x^j_t$ gives the ratio of Lyapunov functions at 
two successive times,
\begin{equation}
\frac{V^{ij}_{t+1}}{V^{ij}_t} = (1-\epsilon)^2 \left[ f^{\prime}(x^j_t) +
\frac{x^i_t - x^j_t}{2} f^{\prime\prime}(x^j_t)+ {\cal O}((x^i_t - x^j_t)^2)
\right]^2 \nonumber
\end{equation}
If the term in the square bracket on the RHS is bounded then there
will be a critical value of $\epsilon$ beyond which
$\frac{V^{ij}_{t+1}}{V^{ij}_t} < 1$ and thus the bipartite synchronized 
state will be stable.

We see that for d-synchronization, Lyapunov function for any pair
of nodes does not depend on the size of 
the complete bipartite network and type
of the coupling because in the expression for Lyapunov function,
contribution of such couplings cancel out. This is not the case for
globally coupled networks where contribution of the coupling terms for 
the two nodes under consideration do not exactly cancel and the size of the
network has an effect on the asymptotic behaviour. 

For the logistic map and using $0 \leq x^i_t + x^j_t \leq 2$
and Eq. (\ref{cond-syn}), we get the following range for $\epsilon$ values
for synchronization of nodes $i$ and $j$,
\begin{equation}
\frac{\mu-1}{\mu} \leq \epsilon \leq 1
\end{equation}
A better $\epsilon$ range can be obtained by taking a more appropriate  
boundary as $X^- < x^i_t+x^j_t < X^+$, which gives
\begin{equation}
1- \frac{1}{\mu A} < \epsilon \leq 1 
\nonumber
\end{equation}
where $A$ is defined after Eq.~(\ref{range-gcm-lyafun-fx}).

\subsection{Journey from small network to large network}
To understand the features of d-synchronization in bipartite network
and comparison of this large network with the network of three nodes, 
here we briefly present the numerical results for logistic map 
with $\mu=4$. Fig.~\ref{lya-3-fx}(b) plots the largest Lyapunov exponent
as function of $\epsilon$ 
and Fig.~\ref{inter-intra-3-fx}(b) plots $f_{inter}$ and $f_{intra}$ as
a function of $\epsilon$ for a bipartite network with $g(x)=f(x)$ and $\mu=4$. 
Comparing with the three nodes network (Figs.~\ref{lya-3-fx}(a)
and~\ref{inter-intra-3-fx}(a)) we find that the major difference is
in the range $0.20.. <\epsilon < 0.5$ where the turbulent region reappears
for the three nodes network while the mixed region having both d- and
s-clusters is observed for the larger bipartite networks.
The region boundaries for the  period two, globally s-state
and fixed point are the same for the three nodes and larger bipartite 
networks and are the same as given for cases II, III and I respectively for 
three nodes bipartite network with $g(x)=f(x)$. 
We see that most of the dynamical features and synchronized cluster
formation in coupled
map bipartite networks with three nodes get carried over to the 
larger bipartite networks.

As in the case $g(x)=f(x)$ there is a similarity between the three nodes 
bipartite network and the larger bipartite networks for $g(x)=x$.  
Fig.~\ref{lya-3-x}(b) plots the largest Lyapunov exponent as a function
of $\epsilon$
and Fig.~\ref{inter-intra-3-x}(b) plots $f_{inter}$ and $f_{intra}$ as
a function of $\epsilon$ for a bipartite network with $g(x)=x$ and $\mu=4$.
Comparing with the three nodes network (Figs.~\ref{lya-3-x}(a)
and~\ref{inter-intra-3-x}(a)) we find that the main difference is
in the range $0.24.. <\epsilon < 0.35..$ where the turbulent region reappears
for the three nodes network while d-synchronization appears for
the larger bipartite networks. The other regions show a very similar
behaviour for both three nodes and larger bipartite networks.

\subsection{S- and D-synchronization}
The analysis presented so far
shades some light on the dynamical origin of the two types
of synchronization namely self-organized and driven, that we have
studied. In s-synchronization the clusters have dominant
intra-cluster couplings while in d-synchronization they have
dominant inter-cluster couplings. We consider two nodes and globally
coupled networks as simple examples of s-synchronization 
while three nodes and complete bipartite networks as simple examples of 
d-synchronization. The dynamics of the difference variable and
Lyapunov function analysis are useful for understanding
the difference between the two mechanisms of cluster formation.

From Eq.~(\ref{add-diff-2})(b) we see that in the dynamics of the difference
variable $d_t$ for the two nodes network the coupling term adds an extra decay term.
This is also seen from
the expression (\ref{lyafun-2}) for Lyapunov function.
On the other hand, from Eq.~(\ref{diff-3}) we see that in the
dynamics of the difference variable for the three nodes network the coupling  
terms with the third variable cancel out (see also Eq.~(\ref{Lyafun-3})
for Lyapunov function). The situation is more complicated when we consider
larger networks. The d-synchronization shows the same trend i.e.
cancellation of the coupling terms in the dynamics of the difference
variables and as well as in the expression for Lyapunov function 
(Eq.~(\ref{Lyafun-3})).
On the other hand, Eq.~(\ref{lyafun-global}) for Lyapunov function for
the globally coupled maps shows that
the direct coupling term between the two nodes under consideration
adds an extra term in the difference variable while the coupling terms to other
variables cancel out.

The analysis presented in this paper is for exact synchronization while
Ref.~\cite{sarika-REA2} considers phase synchronized clusters. 
However, we feel that the
dynamical origin for the two mechanisms of cluster formation should
be similar in both cases. This is supported by the very similar features of
plots of phase space, largest Lyapunov exponent, $f_{inter}$ and $f_{intra}$
for the three nodes and complete bipartite networks and the corresponding
plots for several networks considered in Ref.~\cite{sarika-REA2}.

\section{Conclusion}
We have studied s- and d-synchronization in coupled
map networks using some simple networks, namely two and three node
networks and their natural generalization to globally coupled and
bipartite networks respectively. For this study we use both linear stability
analysis and Lyapunov function approach. 
We see that for the difference variable for any two nodes that are in
d-synchronization all the coupling
terms cancel out whereas when they are in s-synchronization
though coupling terms for couplings to other nodes may cancel,
the coupling terms corresponding to the direct coupling between the
two nodes under consideration do not cancel.

The phase diagrams for three nodes network has features very similar to
the different kinds of networks studied in
Ref.~\cite{sarika-REA2}. (The phase diagrams
for two nodes, not shown in this paper, have similar
features except the 
d-synchronization which needs at least three nodes to manifest itself.)
Under the time evolution, the coupled dynamics lies
on  different periodic or chaotic attractors with varying coupling
strength values. The type of coupling function plays an important role
in the time evolution of the coupled dynamics.
We also see that most of the features of coupled dynamics of
small networks with two or three nodes, are carried over to the larger networks
of the same type. 

This work was supported in part by the National Science
Council of the Republic of China (Taiwan) under Grant No.
NSC 91-2112-M-001-056.


\begin{thebibliography}{99}
\bibitem{note1} In this case the increased range of stability with
increasing $N$ is accompanied by a decrease in the basin of
attraction. Hence, numerically starting from random initial conditions 
this incremental part of the range is not observed.
\end{thebibliography}
\end{document}